\documentclass[twocolumn,pra,nofootinbib, superscriptaddress,showpacs]{revtex4-1}
\usepackage{amssymb}
\usepackage{amsmath}
\usepackage{graphicx}
\usepackage{mathptmx}
\usepackage{epsf}
\usepackage{tikz}
\usepackage{epsfig}
\usepackage{epstopdf}
\usepackage{color}
\usepackage{color}
\usepackage{float}
\usepackage[capitalise]{cleveref}
\usepackage{comment}

\def\g{{\bf{g}}}

\def\g0{{\gamma_0}}

\newcommand{\llangle}{{\langle\!\langle}}
\newcommand{\rrangle}{{\rangle\!\rangle}}
\newcommand{\tr}{{\rm Tr}}

\begin{document}
\newpage
\null
\newpage
\title{Dynamics of a Single Spin in Markovian and non-Markovian Environments}
\title{Spin-boson model under dephasing: Markovian vs Non-Markovian dynamics}
\author{Naushad Ahmad Kamar}
\author{Daniel A. Paz}
\author{Mohammad F. Maghrebi}
\affiliation{Department of Physics and Astronomy, Michigan State University, East Lansing, MI 48823, USA}
\begin{abstract}
The spin-boson model, describing a two-level system strongly coupled to a bosonic bath, is extensively studied as a paradigmatic dissipative quantum system, exhibiting rich dynamical behavior and even a localization transition in the strong coupling regime. Here, we additionally consider dephasing as a source of Markovian dissipation on top of the non-Markovian dynamics due to an Ohmic bath, and investigate the dynamics of the spin.  We show that the characteristic frequency of the spin dynamics, while strongly renormalized by the bosonic bath, changes in a simple fashion (or doesn't change at all) with dephasing.  
To obtain these results, we develop an exact non-perturbative method known as the stochastic Schr\"{o}dinger equation, mimicking the Ohmic bath via a stochastic magnetic field combined with the Lindblad quantum master equation due to dephasing, which allows us to numerically compute the dynamics.
Furthermore, we derive weak-coupling analytic results utilizing the well-known non-interacting blip approximation. 
Our findings are relevant to quantum simulation of the spin-boson model in the regime of strong coupling in trapped ions and circuit QED architectures among others.
\end{abstract}
\maketitle
\section{Introduction}
A quantum impurity coupled to a bath defines a paradigmatic problem in quantum many-body physics. It leads to emergent phenomena ranging from the Anderson orthogonality catastrophe \cite{Anderson_Infrared_Catastrophe}, and the X-ray edge problem \cite{nozieres_X_Ray_edge}, to the resistivity upturn in the Kondo problem \cite{kondo_effect}. Furthermore, the paradigm of the spin-boson model provides a powerful computational approach to strongly correlated many-body systems via dynamical mean field theory \cite{georges_DMFT}.
In general, coupling to the surrounding environment entangles the impurity with the degrees of freedom in the environment, and leads to dissipative dynamics.  
Maintaining the coherence in qubits in spite of the coupling to environment is a fundamental challenge in quantum computation and simulation.

A widely studied quantum impurity problem is the so-called spin-boson model (also intimately related to the Kondo physics) where a two-level spin is coupled to a bath consisting of many bosonic degrees of freedom usually considered as an infinite collection of harmonic oscillators \cite{Leggett_spin_boson_model}. 
The coupling between the spin and the bath can be fully characterized by the bath spectral function $J(\omega)$. For an Ohmic bath characterized by $J(\omega)\sim \alpha \omega$, the spin-boson model exhibits distinct phenomena depending on the coupling $\alpha$ between the spin and the bath such as (underdamped) coherent oscillations ($0<\alpha<1/2$), incoherent damping ($1/2<\alpha<1$), and a delocalized-to-localized quantum phase transition ($\alpha>1$) \cite{Leggett_spin_boson_model}. A characteristic feature of the spin-boson model  is the strong renormalization of the underdamped oscillations due to the coupling to the Ohmic bath when $0<\alpha<1/2$. 
The strong coupling regime in the spin-boson model has been recently realized in superconducting quantum circuits \cite{Diaz_Ultra_Strong_Spin_Boson_model,magazzu2018probing}. 

While the spin-boson model was originally introduced in the domain of condensed matter physics, there are various proposals realizing this model in quantum simulation platforms. In particular, ultrastrong coupling of an artificial atom to an  electromagnetic continuum---mimicking the bath---has been recently observed in superconducting circuits \cite{Diaz_Ultra_Strong_Spin_Boson_model, yoshihara2017superconducting, mirhosseini2019cavity}. Beside superconducting qubits \cite{Diaz_Ultra_Strong_Spin_Boson_model,LEHUR2016808}, trapped ions \cite{Porras2008,Lemmer_2018} and cold atoms \cite{Dima_spin_in_bath} have also emerged as versatile platforms for realizing the spin-boson model; more generally, models where one or many spins are coupled to a single or several bosonic modes have been realized or proposed in a wide range of platforms \cite{niemczyk2010circuit,Ritsch2013,Leroux2010, britton2012engineered, Kim2009, Porras2004,Sorensen1999}. 
Quantum simulation in many such platforms rely on driving the system in order to engineer an effective Hamiltonian in the \textit{rotating} frame. A regime of immense interest is  where an (ultra)strong coupling between a two-level system and the bosonic environment is achieved. 

Quantum simulation thus provides an  attractive alternative for exploring quantum impurity problems \cite{frisk2019ultrastrong}. However, \textit{unwanted} dissipation, for example the noise in lasers, cannot be avoided in these platforms \cite{Burger_2022}.  This unwanted feature should be contrasted with the desired dissipation due to the coupling to the bosonic bath: the former may be approximated as Markovian and typically results from weak coupling to an environment, while the latter is desired in the regime of strong coupling, and is therefore non-Markovian by nature.
A timely question is then how the quantum characteristics---from coherent oscillations to a localization transition---of the spin-boson model are affected in the presence of the unavoidable Markovian dissipation. 
And, how should one describe the competition between Markovian and non-Markovian dynamics? 
A challenge presents itself immediately: in the presence of the drive and Markovian dissipation in quantum simulation platforms, the resulting spin-boson model is inherently driven-dissipative. That is, the system will not be in its ground state even if the bosonic modes are at zero temperature, but will instead approach a non-equilibrium steady state as the result of the competition between drive, dissipation, and the coupling to the bosonic bath. These questions have been investigated recently in mean-field Dicke-type models \cite{Nagy_2015,Nagy_2016,Lundgren_2020}; however, a strongly interacting spin-boson model poses a formidable challenge. To this end, a relatively large toolbox has been developed to tackle this problem including Bethe ansatz \cite{Ponomarenko_1993}, functional-integral approaches \cite{Leggett_spin_boson_model}, renormalization group \cite{Shapourian_2016}, a stochastic Schr\"{o}dinger-like equation (SSE) \cite{Peter_Stochastic_Schrodinger_Equation}, the widely used \textit{non-interacting blip approximation} (NIBA) \cite{Dekker_NIBA,Leggett_spin_boson_model}, and more recently tensor-network methods \cite{Alex_Chin_Spin_Boson2016,Plenio_Spin_Boson2010,Wall_2016,Keeling_MPDO} among other things.

In this paper, we investigate the dynamics of the spin-boson model coupled to an Ohmic bath with $0<\alpha<1/2$ while subject to Markovian dissipation. Specifically, we investigate how the underdamped oscillations
and their characteristic renormalized frequency are affected in the presence of the Markovian bath. We  consider dephasing along different axes as the primary source of Markovian loss. For dephasing
along the axis of the spin-only Hamiltonian (decoupled from bosonic modes), 
we find that the characteristic frequency is barely dependent on Markovian dissipation, underscoring its robustness against dissipation. For dephasing along the axis set by coupling to the bosonic bath, we find that the frequency decreases with dissipation in a simple fashion, and that the dynamics becomes overdamped at large values of dephasing.  

We obtain the above results by a combination of a numerically exact method based on a nonperturbative SSE as well as the widely-used NIBA \cite{Leggett_spin_boson_model,Dekker_NIBA}. 
The former method is based on a reparametrization of the spin configuration in the path integral combined with the influence functional 
of the spin-bath coupled system \cite{Peter_Stochastic_Schrodinger_Equation}. 
We simplify, adapt and extend the method pioneered in Ref.~\cite{Peter_Stochastic_Schrodinger_Equation} to the case of Markovian (on top of non-Markovian) dissipation.
As a complementary approach, we derive analytic results based on NIBA which provides a weak-coupling approximation in the context of the spin-boson model. 

The structure of this paper is as follows. In \cref{sec:Model}, we introduce the spin-boson model in the presence of the Markovian dissipation. We derive the SSE for the dynamics in \cref{sec:SSE}, and provide the numerically exact results in \cref{sec:Numerics}. In  \cref{sec:NIBA}, we provide the analytical results from the NIBA 
in the presence of Markovian dissipation, and show excellent agreement with the numerically exact results. We also discuss and interpret our results in this section. Finally in \cref{sec:Outlook}, we summarize our findings and remark on interesting future directions. We present our simplified derivation of the SSE in \cref{sec:stochastic_sch}, and a detailed derivation of the NIBA in \cref{App:NIBA}.
\section{Spin-boson model under Markovian loss}\label{sec:Model}
In this section, we introduce the main model. Let us first consider the paradigmatic spin-boson model describing a two-level system S coupled to an infinite number of non-interacting bosons denoted by B. The system-bath model is described by the Hamiltonian ${H}$:
\begin{equation} \label{eq:eq0a}
\begin{split}
{H}=\frac{\Delta}{2} \sigma_x+
\sum_k\omega_k b_k^\dagger b_k+\frac{\sigma_z}{2}\sum_k \lambda_k (b_k^\dagger+b_k)\,.
\end{split}
\end{equation}
We denote the three terms on the rhs by $H_S$, $H_B$, and $H_{SB}$ representing the system, the bath and the linear coupling between the  two, respectively. The effective coupling between the spin and the bath depends on $\omega_k$ and $\lambda_k$, and is fully characterized by the spectral function defined as $J(\omega)=\pi\sum_k \lambda_k^2\delta(\omega-\omega_k)$. Specifically, for an Ohmic bath, the spectral function is given by 
\begin{equation} \label{eq:eq0a1}
\begin{split}
J(\omega)=2\pi\alpha \omega e^{-\omega/\omega_c},
\end{split}
\end{equation}
where 
the interaction parameter $\alpha$ controls the properties of the spin and $\omega_c$ is the frequency cutoff of the bath. In this work, we consider an Ohmic bath with $0<\alpha<1/2$. It is well known that, in this regime, the spin exhibits damped oscillations at a frequency
\begin{equation}\label{eq:Delta_ren}
\Delta_r=\Delta\left({\Delta}/{\omega_c}\right)^\frac{\alpha}{1-\alpha},
\end{equation}
which is strongly renormalized by the coupling to the Ohmic bath, and exhibits a universal dependence on $\alpha$, a feature that is intimately related to the Kondo physics \cite{Leggett_spin_boson_model}. Notice that $\Delta_r$ is smaller than the bare value $\Delta$ because spin transitions are suppressed in the presence of a cloud of bath modes \cite{Peter_Stochastic_Schrodinger_Equation}.
We emphasize that the bosonic bath considered above constitutes a non-Markovian bath in general since the coupling is generally of the same order as the energy scales of the system. Moreover, the spectral function changes significantly with frequency, which should be contrasted with that in a Markovian environment which is frequency independent at the relevant (optical) frequencies.

Now motivated by the quantum simulation proposals for the spin-boson model,  
we also consider Markovian dissipation due to environmental sources of noise.
The resulting dynamics is more generally governed by a quantum master equation as
\begin{equation} \label{eq:eq0b1}
\frac{d\rho(t)}{dt}=-i[H,\rho]+\sum_\mu \Big[L_\mu\rho L_\mu^\dagger -\frac{1}{2}(L_\mu^\dagger L_\mu \rho +\rho L_\mu^\dagger L_\mu)\Big],
\end{equation}
where $H$ is the Hamiltonian defined in \cref{eq:eq0a}, and $L_\mu$s describe different types of Lindblad operators characterizing Markovian dissipation. 
In this work, we consider two examples of Markovian dissipation: dephasing $L_{dph}=\sqrt \Gamma_\phi\sigma_z$; and, ``depolarization'' $L_x=\sqrt \Gamma_x\sigma_x$. While depolarization is commonly referred to as a quantum channel where the Block sphere contracts uniformly, here we have used it to refer to dephasing along the $x$ direction.  We also emphasize that $H$ should be interpreted as the Hamiltonian in the rotating frame; for example, see \cite{Porras2008,Lemmer_2018} for drive schemes where the above dynamics is realized in trapped ions. The driven nature of the model is thus disguised in the rotating frame. Put differently, \cref{eq:eq0b1} describes an inherently driven-dissipative system which approaches a non-equilibrium steady state \cite{diehl2008quantum,verstraete2009quantum} even if the bosonic modes in B are at zero temperature. An alternative  perspective is to consider the Hamiltonian $H$ as the native Hamiltonian, while the two (Markovian and non-Markovian) baths are mutually out of equilibrium. This is conceptually similar to a system coupled to two baths at different chemical potentials, resulting in a non-equilibrium steady state \cite{kamenev2023field}. 

In this work, we like to investigate the interplay of coherent dynamics, coupling to the bosonic modes, and Markovian dissipation. In particular, we investigate if and how the renormalized frequency in \cref{eq:Delta_ren} changes in the presence of Markovian dissipation.  
We show that, rather surprisingly, that the frequency is unaffected by depolarization even at moderate values of $\Gamma_x$; in contrast, the frequency decreases in the presence of dephasing, and the dynamics becomes overdamped at sufficiently large values of $\Gamma_\phi$. To this end, we first develop a nonperturbative method to exactly simulate the dynamics. 
\section{Stochastic Schr\"{o}dinger Equation} \label{sec:SSE}
In this section, we introduce a non-perturbative technique for an exact numerical solution of the dynamics via the SSE method. A similar approach has been developed for the spin-boson model in the absence of Markovian bath \cite{Peter_Stochastic_Schrodinger_Equation}; see also \cite{lesovik2002dynamics}. We first provide a brief introduction of this method for the standard spin-boson model before considering Markovian loss.  

A first step is to vectorize the density matrix via $|i \rangle\langle j| \to |ij\rrangle \equiv |i\rangle\otimes |j\rangle$. The Hamiltonian dynamics can be then expressed as
\begin{equation} \label{eq:eq0b}
\begin{split}
|\rho(t)\rrangle=e^{-i(H^u-H^l)t}|\rho(0)\rrangle,
\end{split}
\end{equation}
where we have defined $H^u=H\otimes I$ and $H^l=I\otimes H^T$. 

In a convenient basis where $\sigma_{x,z}$ as well as $b_k$ and $b_k^\dagger$ (in the number basis) are real, we have $H^T= H^*=H$. We further assume that at time $t=0$ the bath is at the inverse temperature $\beta$, and is decoupled from the spin which is initially in a state given by the density matrix $\rho_S(0)$. Hence, the initial density matrix of the system plus bath is $\rho(0)=\rho_S(0)\otimes e^{-\beta H_B}$. In the vectorized form, the initial density matrix is then given by 
\begin{equation} \label{eq:eq0b2}
\begin{split}
|\rho(0)\rrangle&=|\rho_S(0)\rrangle\otimes|e^{-\beta H_B}\rrangle\,.
\end{split}
\end{equation}
We then trace out the bath degrees of freedom to find the reduced density matrix of the spin, $\rho_S(t)=\text{Tr}_B(\rho(t))$. In our vectorized notation, the latter density matrix is given by $|\rho_S\rrangle=\llangle I_B|\rho(t)\rrangle$ where $\llangle I_B|$ is the vectorized form of the identity matrix corresponding to the bath. We can thus write 
\begin{equation} \label{eq:eq0b3}
\begin{split}
|\rho_S(t)\rrangle=\llangle I_B|e^{-i(H^u-H^l)t}|\rho(0)\rrangle\,.
\end{split}
\end{equation}
The bath degrees of freedom can be traced out exactly in an elegant fashion using the Feynman-Vernon formalism \cite{feynman_Influence_Functional, weiss2012quantum}.
The resulting \textit{influence functional} comes with nontrivial kernels that involve a long-range coupling between the spin variables at different times. Assuming an Ohmic bath and $\omega_c\gg \Delta$, the kernel corresponding to the retarded (causal) component becomes local in time, while the kernel corresponding to the quantum fluctuations of the bosonic bath can be dealt with using a Hubbard-Stratonovich transformation. The result is the SSE that can be efficiently simulated in the limit $0<\alpha<1/2$ and large $\omega_c$. We refer the interested reader to \cref{sec:stochastic_sch} for the details, and just quote the expression for the state at time $t$: 
\begin{equation} \label{eq:eq14ab}
\begin{split}
|\rho_S(t)\rrangle&=\prod_{m=0}^{\infty} \int \frac{dx_m}{\sqrt {2\pi}} e^{-x_m^2/2}  {\rm T}_te^{-i\int_0^t ds {\cal A}(s)}|\rho_S(0)\rrangle. 
\end{split}
\end{equation}
Here, ${\rm T}_t$ denotes time ordering, and the matrix ${\cal A}(t)$ is given by (in the $\sigma_z$ basis)
\begin{equation}\label{eq:cal_A}
    \begin{split}
{\cal A}(t)&=
\begin{pmatrix}
0&-\frac{\Delta}{2}&\frac{\Delta}{2} &0\\
-\frac{\Delta}{2} e^{i\pi\alpha}&{h}(t)&0&\frac{\Delta}{2} e^{-i\pi\alpha}\\
\frac{\Delta}{2} e^{-i\pi\alpha}&0& -h(t)&-\frac{\Delta}{2} e^{i\pi\alpha}\\
0& \frac{\Delta}{2}&-\frac{\Delta}{2}&0
\end{pmatrix},
\end{split}
\end{equation}
with the function $h(t)$ defined as
\begin{equation} \label{eq:eq14abc}
\begin{split}
h(t)&=\sum_{m=0}^{\infty} x_m \sqrt{\frac{G_m}{\pi}}\psi_m(t)\,.
\end{split}
\end{equation}
Here, $G_m$ and $\psi_m(t)$ are known variables and functions defined in \cref{sec:stochastic_sch}. Notice that $h(t)$ mimics a \textit{stochastic} longitudinal field  as the coefficients $x_m$ are drawn from a normal distribution. The expression in \cref{eq:eq14ab} can be computed by solving a time-dependent Schr\"{o}dinger equation as 
\begin{equation}\label{eq:eq17a}
\frac{d}{dt}|\psi(t)\rangle = -i {\cal A}(t) |\psi(t)\rangle,
\end{equation}
where $|\psi\rangle$ represents the state of a four-level system.
 
 In practice, the integral over $x_m$ is performed by sampling from a normal distribution \cite{Peter_Stochastic_Schrodinger_Equation}. For each realization, we generate $x_0, x_1,..., x_{m_{max}}$, which we may truncate at the order $m_{max}$, compute $h(t)$ defined in Eq.~(\ref{eq:eq14abc}), which is then substituted in \cref{eq:eq17a} to solve for $|\psi(t)\rangle$ as a function of time. Finally, we take the arithmetic average of $|\psi(t)\rangle$ over different realizations which yields the vectorized density matrix $|\rho_S(t)\rrangle=\overline{|\psi(t)\rangle}$ with the bar indicating the average over different realizations.  

Solving the SSE allows us to compute quantities of interest; for example, $\langle \uparrow|\rho_S(t)|\uparrow\rangle=\llangle \uparrow\uparrow| \rho_S\rrangle = \overline{\psi_1(t)}$, is given by the first component of the vector $|\psi(t)\rangle$ upon averaging over different realizations. 
We will be particularly interested in the expectation value of  $\sigma_z$ which is given by 
 \begin{equation} \label{eq:eq17ab}
\begin{split}
\langle\sigma_z(t)\rangle&=2 \overline{\psi_1(t)}-1.
\end{split}
\end{equation}
Finally, we remark that our derivation of the SSE (see \cref{sec:stochastic_sch}) is rather simple compared to the more involved approach in the literature \cite{Peter_Stochastic_Schrodinger_Equation}. 
\begin{figure}
\begin{center}
\includegraphics[ scale=0.40]
{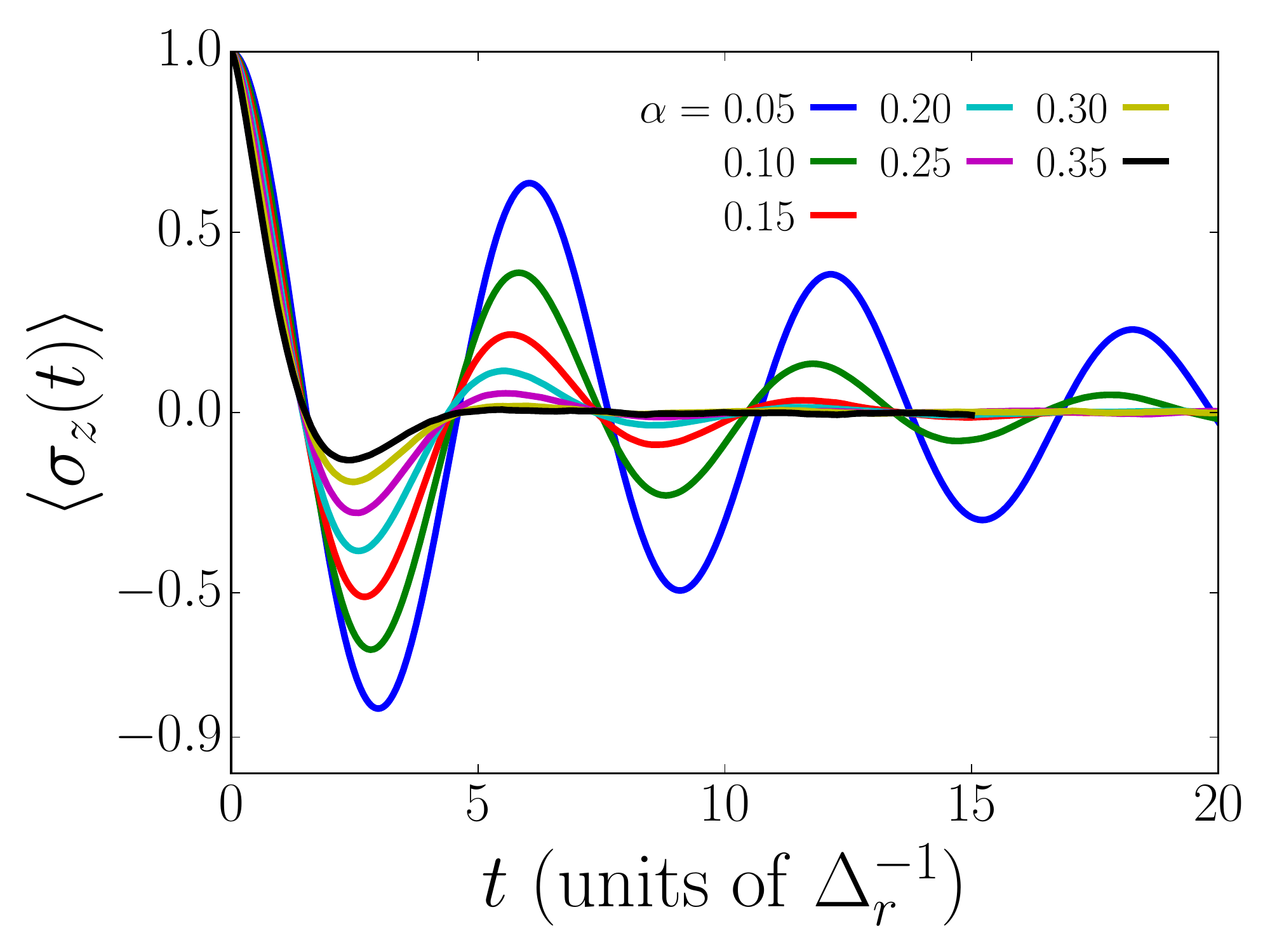}
\hspace*{1cm}\includegraphics [ scale=0.35]
{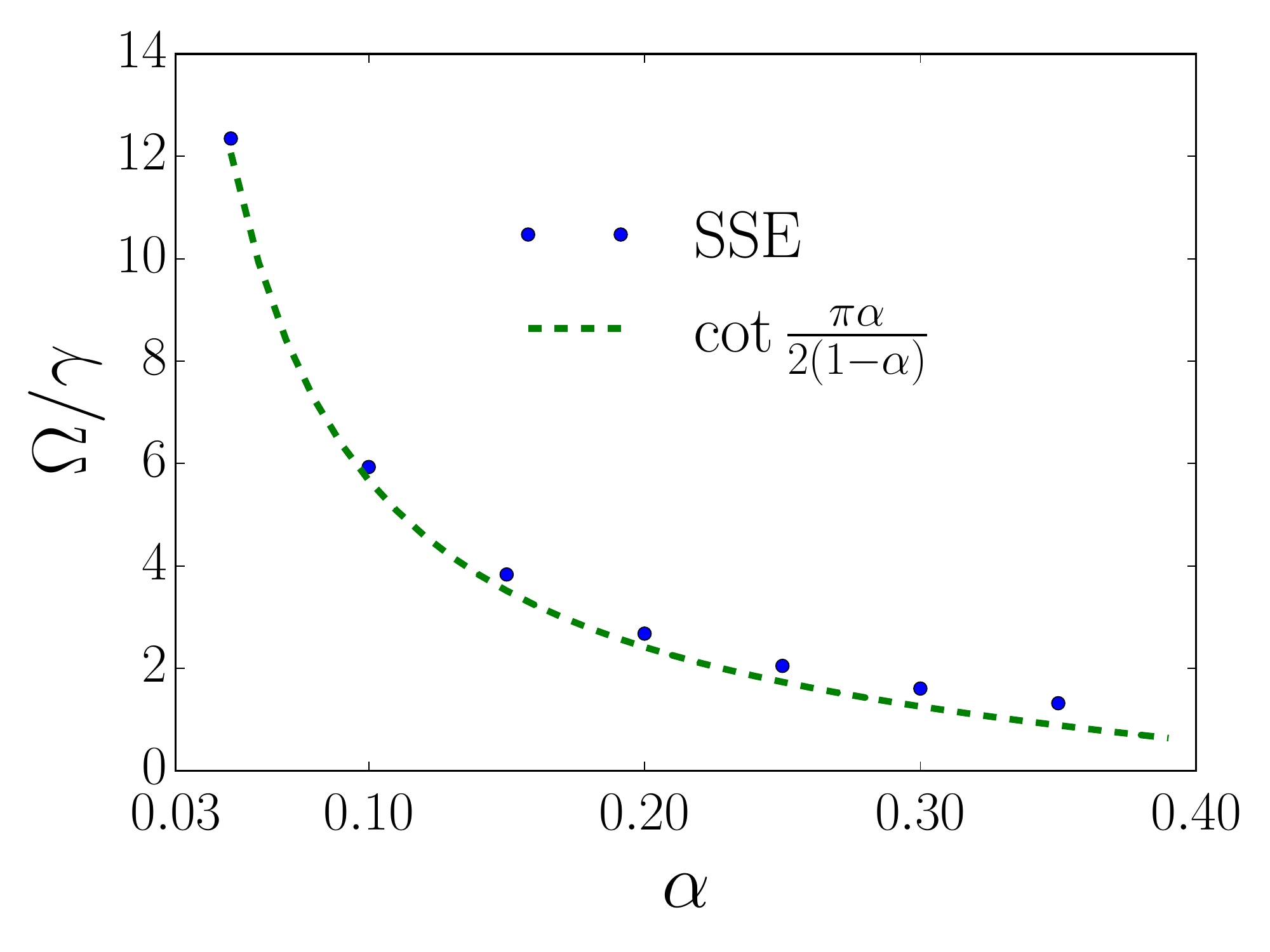}
\end{center} 
\caption{\label{fig:magnetization} 
(color online) Top panel: Magnetization $\langle\sigma_z(t)\rangle$ as a function of time in units of $\Delta_r^{-1}$ without Markovian dissipation at different $\alpha$ using SSE method.
Bottom panel: Quality factor as a function of $\alpha$ using the SSE. We compare the SSE results against the exact relation (dashed line).
Parameters are $m_{max}= 50000$, $\omega_c=100$, and $\Delta=2.0$.}
\end{figure}
\begin{figure}
\begin{center}
\includegraphics [ scale=0.40]
{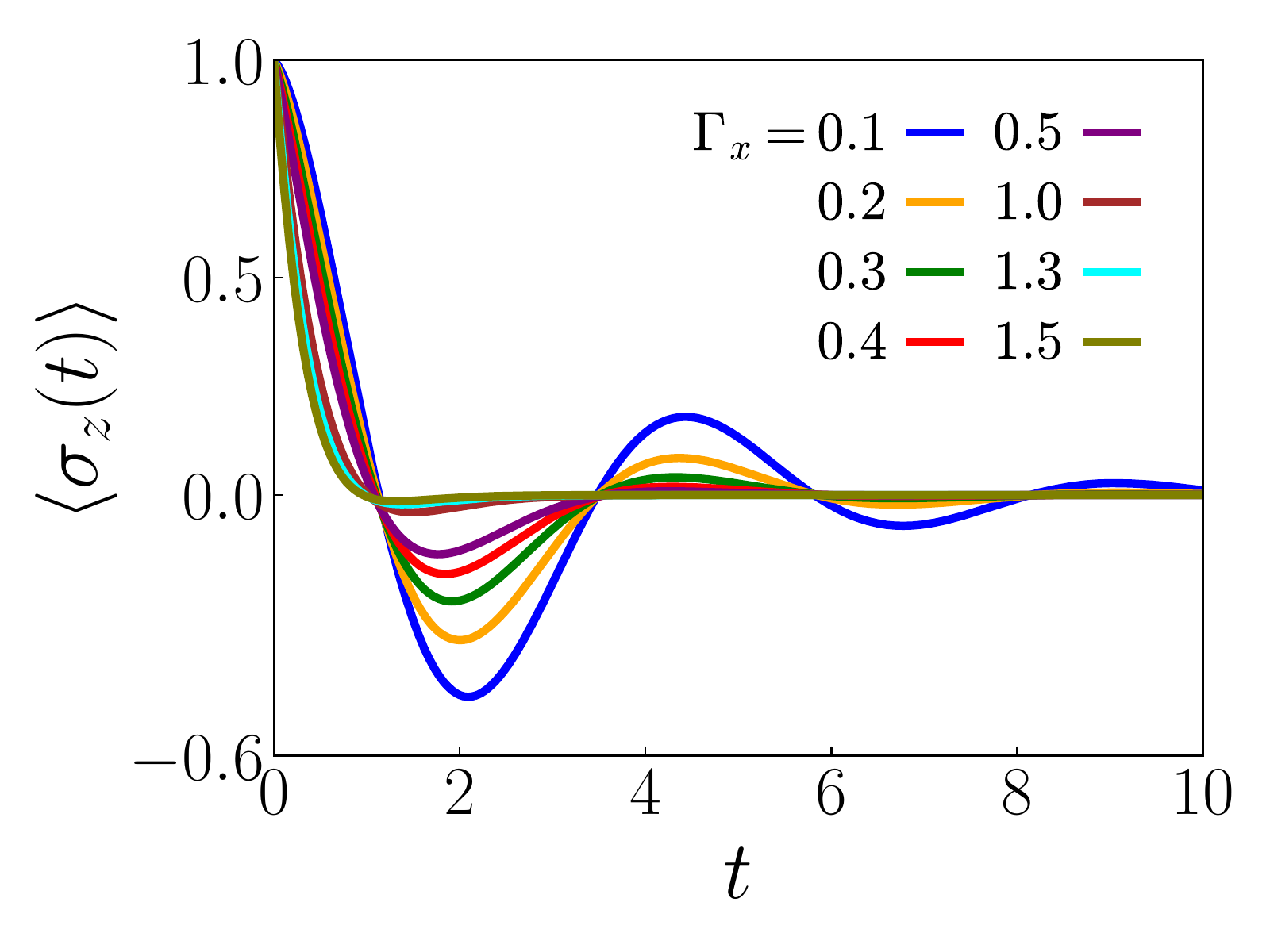}
\includegraphics [scale=0.40]
{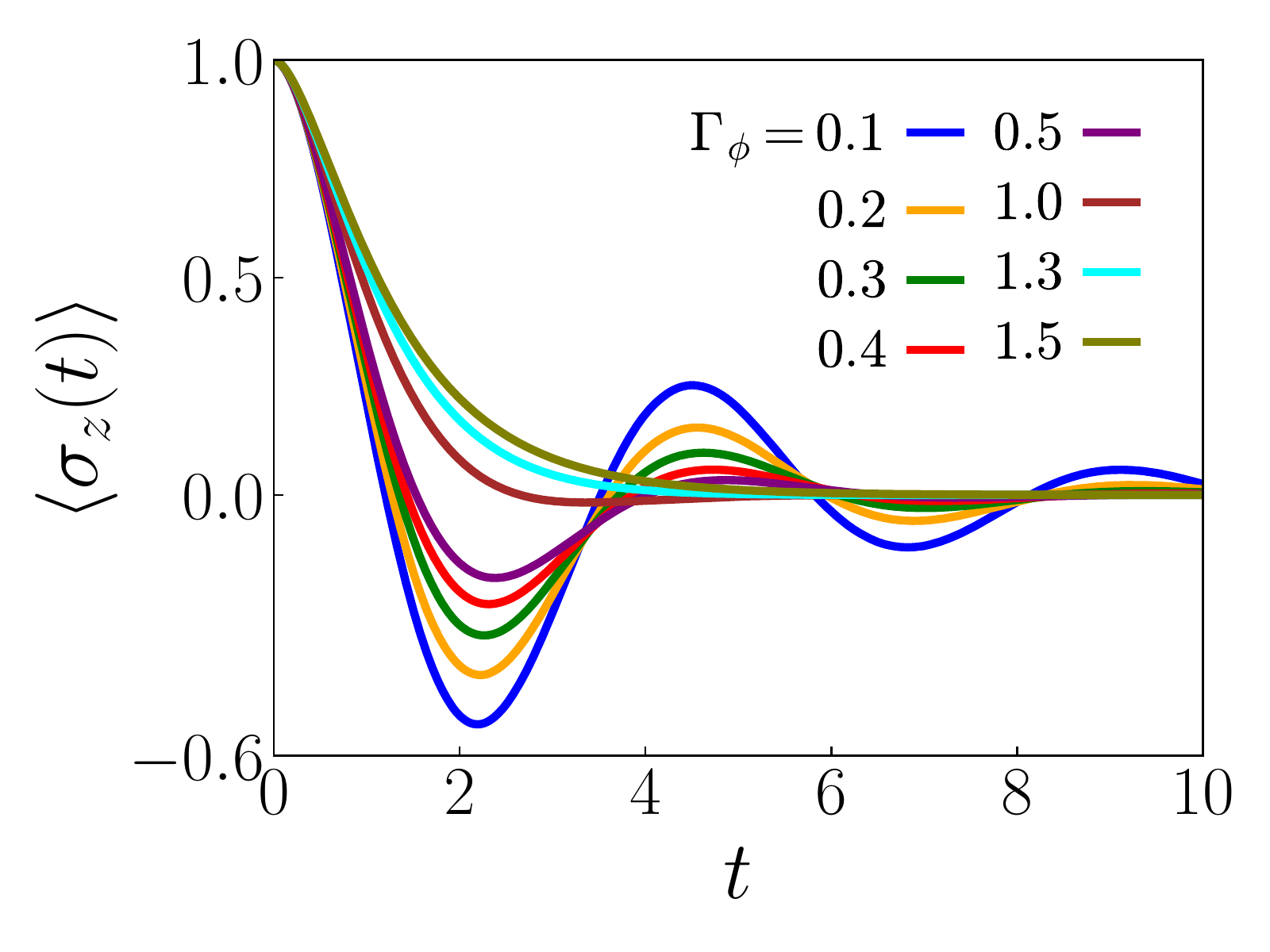}
\end{center} 
\caption{\label{fig:magnetization_SSE_Markovian} (color online). Magnetization $\langle\sigma_z(t)\rangle$ as function of time in the presence of Markovian dissipation using the SSE method; we have set $\alpha=0.1$. Top and bottom panels represent $\langle\sigma_z(t)\rangle$  in the presence of depolarization and dephasing, respectively. Markovian dissipation leads to a faster decay in both cases, although the frequency remains unchanged under  depolarization only.}
\end{figure}
Next, we consider Markovian dissipation which can be naturally and simply incorporated into the SSE. 
Let us first recall that the evolution of the density matrix of the systems plus bath in the presence of Markovian dissipation is governed by the Lindblad master equation in \cref{eq:eq0b1}.
A first step then is to vectorize this equation as
\begin{align} \label{eq:eq15}
\begin{split}
&\frac{d}{dt}|\rho(t)\rrangle=\mathcal{L}|\rho(t)\rrangle,\\
&\mbox{where}\qquad \mathcal{L}=-i(H\otimes I-I\otimes H^{T})  \\
&\qquad +\sum_\mu [L_\mu\otimes L^\ast_\mu -\frac{1}{2}(L^\dagger_\mu L_\mu \otimes I+I\otimes  L^{T}_\mu L^\ast_\mu) ],
\end{split}
\end{align}
where we consider dephasing $L_{dph}=\sqrt \Gamma_\phi\sigma_z$ as well as depolarization $L_{x}=\sqrt \Gamma_x\sigma_x$.
The Hamiltonian involves the bosonic bath which should be traced out systematically. On the other hand, the Markovian dissipation in the last line of the above equation is simply a superoperator that acts only on the spin. Therefore, the same steps leading to the SSE in \cref{eq:eq17a} can be adapted to the full dynamics simply by adding the dissipative superoperator to the matrix ${\cal A}$. The result is a stochastic Schr\"{o}dinger equation  governed by the evolution operator 
\begin{align} \label{eq:eq17d}
&{\cal B}(t)= {\cal A}(t) + \\
&\begin{pmatrix}
-i\Gamma_x& 0 & 0 & i\Gamma_x\\
0 & -i(\Gamma_x+2\Gamma_\phi)& i\Gamma_x&0\\
0 &i\Gamma_x& -i(\Gamma_x+2\Gamma_\phi)&0\\
i\Gamma_x& 0&0& -i\Gamma_x
\end{pmatrix}.\nonumber
\end{align}
A sum over different realizations of the stochastic field is conducted to compute expectations values of observables of interest. With the Ising symmetry, both the model and the formalism bear resemblance to a driven-dissipative quantum Ising model that have been studied recently using a quantum-to-classical mapping \cite{Paz_2021_exact}.
\section{Numerical Results}\label{sec:Numerics}
In this section, we use the SSE to numerically simulate the dynamics of a spin coupled to an Ohmic bath as well as a Markovian bath. To benchmark our method, we first consider the spin dynamics in the absence of Markovian dissipation.  
In particular, we verify that the spin exhibits underdamped oscillations at the renormalized frequency given by \cref{eq:Delta_ren}. 
In order to compute the dynamics using the SSE approach, we take $m_{max}= 10000, 50000$ realizations of the stochastic field for dephasing and depolarization, respectively. 
We start from the initial state $|\uparrow\rangle$ (in the $\sigma_z$ basis),  take the tunneling rate $\Delta=2$ and a large cutoff $\omega_c=100$ where the SSE is applicable. The expectation value $\langle \sigma_z(t)\rangle$ is then computed from \cref{eq:eq17a,eq:cal_A,eq:eq17ab}; we use the fourth-order Runge-Kutta method to solve the time-dependent Schr\"{o}dinger equation.
In \cref{fig:magnetization}, we show $\langle\sigma_z(t)\rangle$ in the absence of Markovian dissipation. 
In the top panel of \cref{fig:magnetization}, we find underdamped oscillations for different values of $\alpha=0.05-0.35$ and show that the frequency of oscillations is the same in units of $\Delta_r$.
We also consider the quality factor $\Omega/\gamma$ where $\Omega$ ($\gamma$) is the frequency (decay rate) of the spin. In the lower panel of \cref{fig:magnetization}, we contrast this factor computed from the SSE against the exact result (obtained from conformal field theory \cite{CFT_Spin_Boson_Model}), and find an excellent agreement.

We now switch on the Markovian dissipation. 
In Fig.~\ref{fig:magnetization_SSE_Markovian}, we show $\langle\sigma_z(t)\rangle$ as a function of time for both depolarization (upper panel) and dephasing (lower panel). 
We make the following observations. For depolarization (i.e., dephasing along the $x$ direction), the dynamics decays faster, but interestingly the frequency is barely dependent on the dissipation strength and  is just set by the (non-Markovian) Ohmic bath. This can also be viewed as a kind of robustness against depolarization. This behavior suggests a non-renormalization of the frequency by the Markovian dissipation. 
On the other hand, the dephasing channel  not only changes the decay rate, but clearly changes the frequency of oscillations as well. In fact, for sufficiently large $\Gamma_\phi$, we find a transition into overdamped dynamics. 

To gain a better analytical understanding of the dynamics in the presence of Markovian dissipation, 
we perform a weak-coupling perturbative approach, widely known as the NIBA in the next section.
\begin{figure}[t]
\begin{center}
\includegraphics [ scale=0.42]
{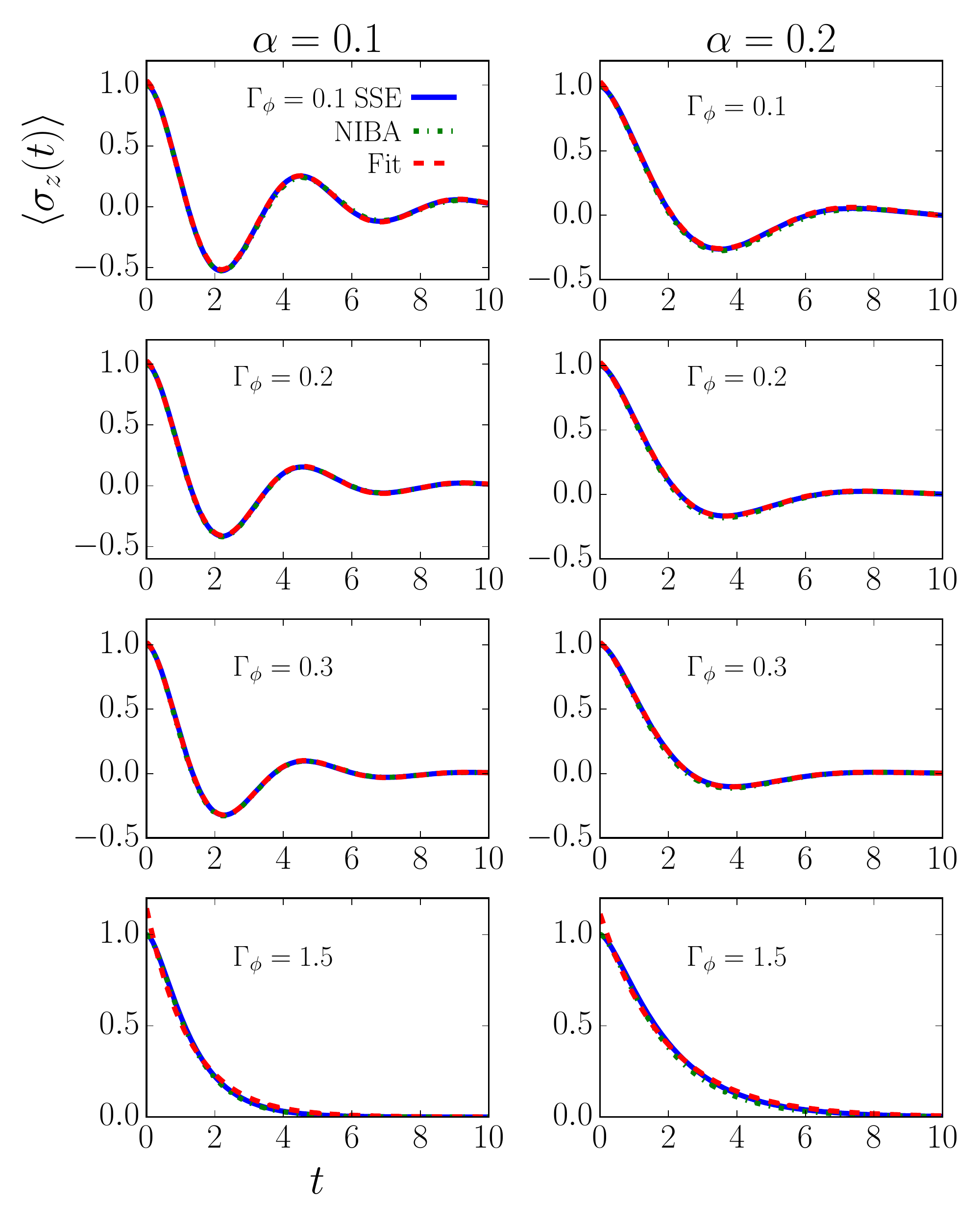}
\end{center} 
\caption{\label{fig:magnetization_depolarization} (color online) Magnetization $\langle\sigma_z(t)\rangle$ as a function of time in the presence of dephasing. The left and right  panels represent  $\alpha=0.1$ and $\alpha=0.2$, respectively. Solid (blue), dotted (green), and dashed (red) lines are computed using SSE, NIBA and a fit of the SSE results to $\langle\sigma_z(t)\rangle=A_0\cos(\Omega t+\phi_0)e^{-\gamma t}$. We find SSE in excellent agreement with the NIBA and the fit to underdamped dynamics.}
\end{figure}
\begin{figure}[t]
\begin{center}
\includegraphics [ scale=0.42]
{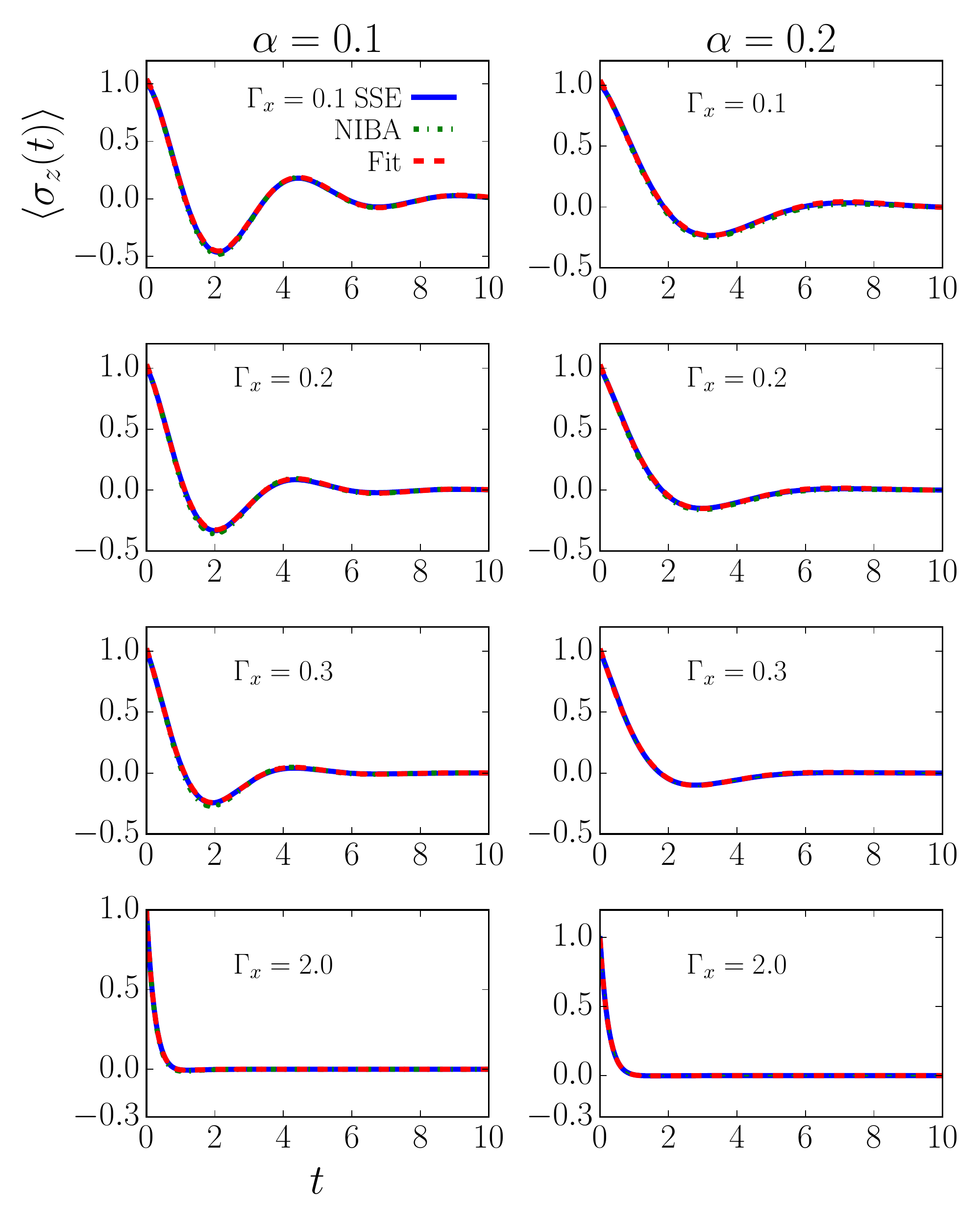}
\end{center} 
\caption{\label{fig:magnetization_dephasing} (color online) Magnetization $\langle\sigma_z(t)\rangle$ as function of time in the presence of depolarization. The left and right panels represent $\alpha=0.1$ and $\alpha=0.2$, respectively. Solid (blue), dotted (green), and dashed (red) lines are computed using SSE, NIBA and a fit of the SSE results to $\langle\sigma_z(t)\rangle=A_0\cos(\Omega t+\phi_0)e^{-\gamma t}$. We find SSE in excellent agreement with the NIBA and the fit to underdamped dynamics.}
\end{figure}
\section{Non-interacting blip approximation}\label{sec:NIBA}
In this section, we compute $\langle\sigma_z(t)\rangle$ for different types of Markovian dissipation using NIBA, which is effectively a weak-coupling approximation that correctly produces the renormalized frequency and quality factor of the spin-boson model for an Ohmic bath \cite{Dekker_NIBA}. In this section, we employ and extend the methodology of NIBA  to the spin-boson model in the presence of Markovian dissipation. To illustrate this method, we first consider the spin-boson model with the spin subject to dephasing. 
The dynamics in our vectorized notation is given by 
\begin{equation} \label{eq:eq0blip1}
\frac{d}{dt}|\rho(t)\rrangle={\cal L}|\rho(t)\rrangle,
\end{equation}
where the Liouvillian superoperator is given by (using $H=H^T$)
\begin{equation} \label{eq:eq1blip1}
{\cal L}=-i[H\otimes I-I\otimes H]+\Gamma_\phi(\sigma_z\otimes \sigma_z-I).
\end{equation}
We then apply a ``polaron'' transformation \cite{Polaron_Transformation_Demler}
via the unitary operator ${\cal U}=\exp(-i\sigma_z B)\otimes \exp(-i\sigma_z B)$ with $B=i\sum_k (\lambda_k/\omega_k)(b_k^\dagger-b_k)$ to obtain\footnote{Upon vectorization, the unitary transformation becomes $U\bullet U^\dagger \to U\otimes U^*$. Since $U\equiv \exp(-i\sigma_z B)$ is purely real, we have ${\cal U} = U\otimes U$.}
\begin{equation} \label{eq:eq2blip1}
\begin{split}
\widetilde{{\cal L}}&=-i[\tilde{H}\otimes I-I\otimes \tilde{H}]+\Gamma_\phi(\sigma_z\otimes\sigma_z-I),
\end{split}
\end{equation}
where $\tilde{H}=\Delta(\sigma_x \cos B-\sigma_y \sin B)/2+\sum_k\omega_k b_k^\dagger b_k$. Notice that the coupling term is now removed but at the expense of the modified form of the first term.
We can then write the following equations of motion for $\sigma_x$, $\sigma_y$, and $\sigma_z$ (in the transformed basis) as
\begin{equation} \label{eq:eq3blip1}
\begin{split}
\frac{d}{dt}\sigma_x&=-\Delta\,\sigma_z \sin B-2\Gamma_\phi \sigma_x,\\
\frac{d}{dt}\sigma_y&=-\Delta \sigma_z \cos  B- 2\Gamma_\phi\sigma_y,\\
\frac{d}{dt}\sigma_z&= \Delta(\sigma_y\cos B+\sigma_x\sin B).\\
 \end{split}
\end{equation}
Next, we solve for $\sigma_z(t)$ exactly and then take the average over the Ohmic bath assuming that
the spin and the bath are uncoupled. We find 
\begin{equation} \label{eq:eq4blip}
\frac{d}{dt}\langle \sigma_z\rangle=-\int_0^t ds e^{-2\Gamma_\phi(t-s)} f(t-s)\langle \sigma_z(s)\rangle\,,
\end{equation}
where $f(t-s)={\Delta^2}/2\langle e^{iB(t)}e^{-iB(s)}+{\rm h.c.}\rangle$. It is convenient to solve this equation using the Laplace transform $f(t) \to  f(\lambda)$ where, in a slight abuse of notation, we use the same symbol for the Laplace-transformed functions. Upon this transformation, we have $f(\lambda)=\Delta_{\rm eff}({\Delta_{\rm eff}}/{\lambda})^{1-2\alpha}$ where we have defined $\Delta_{\rm eff}=[\Gamma(1-2\alpha) \cos(\pi \alpha)]^{1/2(1-\alpha))}\Delta(\Delta/\omega_c)^{\alpha/(1-\alpha)}$ with $\Gamma(x)$ the Gamma function. The solution for the above equation together with the initial condition $\langle\sigma_z(t=0)\rangle=1$ becomes 
\begin{equation} \label{eq:eq5blip}
\begin{split}
\langle \sigma_z(\lambda)\rangle&=\frac{(\lambda+2\Gamma_\phi)^{1-2\alpha}}{\lambda(\lambda+2\Gamma_\phi)^{1-2\alpha}+\Delta_{\rm eff}^{2-2\alpha}}.
 \end{split}
\end{equation}
Taking the inverse Laplace transform of $\langle \sigma_z(t)\rangle$, we then find the dynamics of the spin in real time as 
\begin{equation} \label{eq:eq6blip1}
\begin{split}
\langle\sigma_z(t)\rangle&=\frac{e^{\lambda_1t}(\lambda_1+2\Gamma_\phi)}{2\lambda_1(1-\alpha)+2\Gamma_\phi}+\frac{e^{\lambda_2t}(\lambda_2+2\Gamma_\phi)}{2\lambda_2(1-\alpha)+2\Gamma_\phi}+P_{\rm inc},
 \end{split}
\end{equation}
where $\lambda_{1,2}$ are the two solutions of 
\begin{equation}
\lambda(\lambda+2\Gamma_\phi)^{1-2\alpha}+\Delta_{\rm eff}^{2-2\alpha}=0\,.
\end{equation}
The last term in \cref{eq:eq6blip1} is a contribution from the branch cut in \cref{eq:eq5blip}. In the absence of Markovian dissipation, $P_{\rm inc}$ decays as a power law in time, and becomes dominant at long times, therefore 
NIBA does not give the correct result in this limit.
However, $P_{\rm inc}\sim \exp(-2\Gamma_\phi t)$ decays exponentially under dephasing (see \cref{App:NIBA}), and NIBA correctly captures the qualitative behavior of $\langle\sigma_z(t)\rangle$ 
at all times. 
Therefore, the first two terms in \cref{eq:eq6blip1} determine the nature of the dynamics depending on the eigenvalues $\lambda_{1,2}$. Let us first consider the spin-boson model without Markovian dissipation by setting $\Gamma_\phi=0$. In this case, the two eigenvalues become $\lambda_{1,2}=-\gamma_0\pm i\Omega_0$, resulting in underdamped dynamics with the decay rate $\gamma_0 $ and frequency $\Omega_0$; the subscript 0 denotes the absence of Markovian dissipation. With $\Gamma_\phi=0$, we recover $\Omega_0 =\Delta_{\rm eff} \sin[\pi/(2-2\alpha)]$ consistent with \cref{eq:Delta_ren} for small $\alpha$ \cite{Peter_Stochastic_Schrodinger_Equation}. Turning on the Markovian dissipation, these eigenvalues and the nature of the dynamics could change. In \cref{fig:magnetization_depolarization}, we plot the numerically exact dynamics governed by the SSE for different values of dephasing rate $\Gamma_\phi$ as well as $\alpha$ and contrast the results against the NIBA prediction. We find an excellent agreement between the two. We also observe that the dynamics becomes overdamped at sufficiently large $\Gamma_\phi$. This feature too can be reproduced from the NIBA.

\begin{figure}[ht]
\begin{center}
\includegraphics [ scale=0.165]
{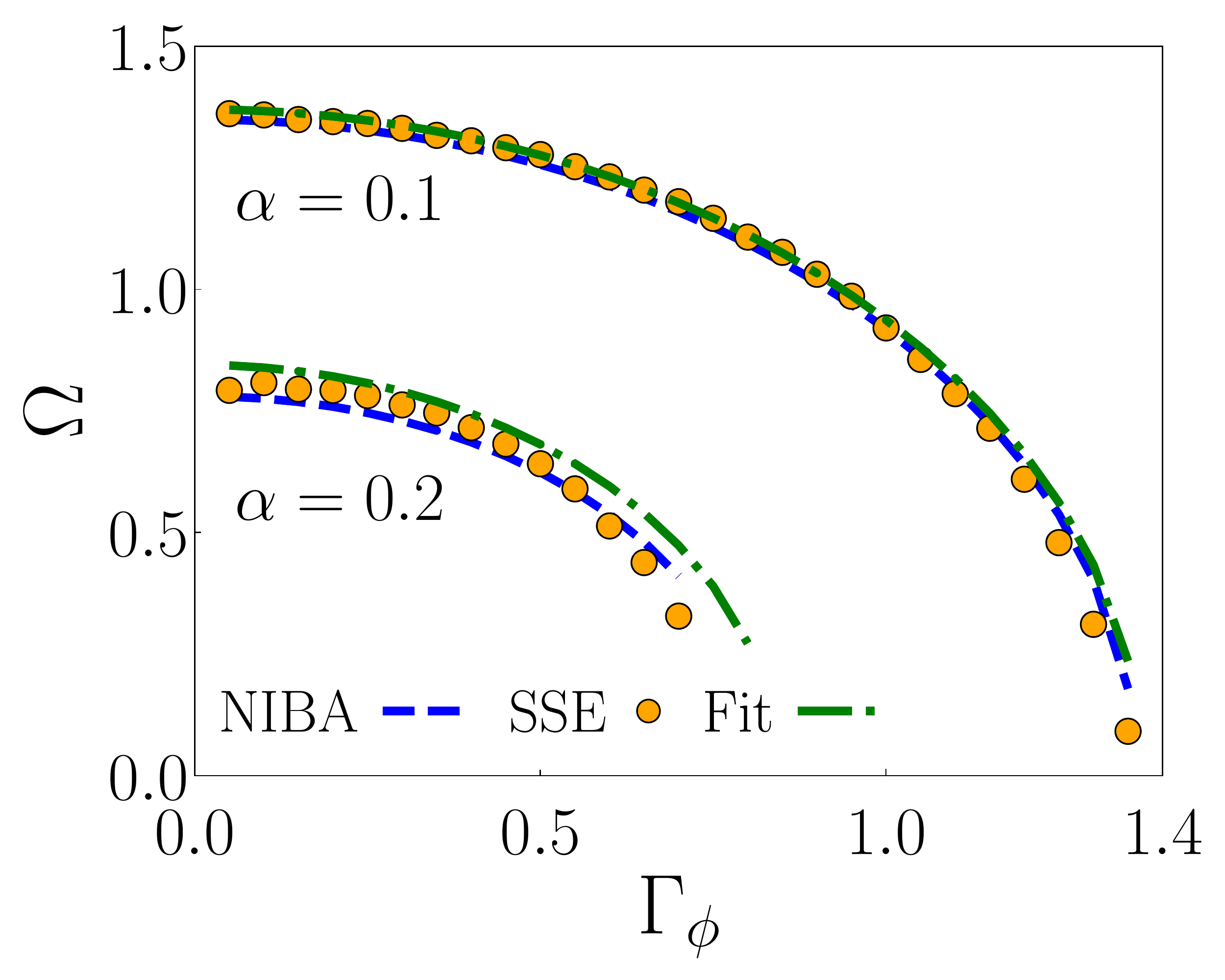}
\includegraphics [ scale=0.165]
{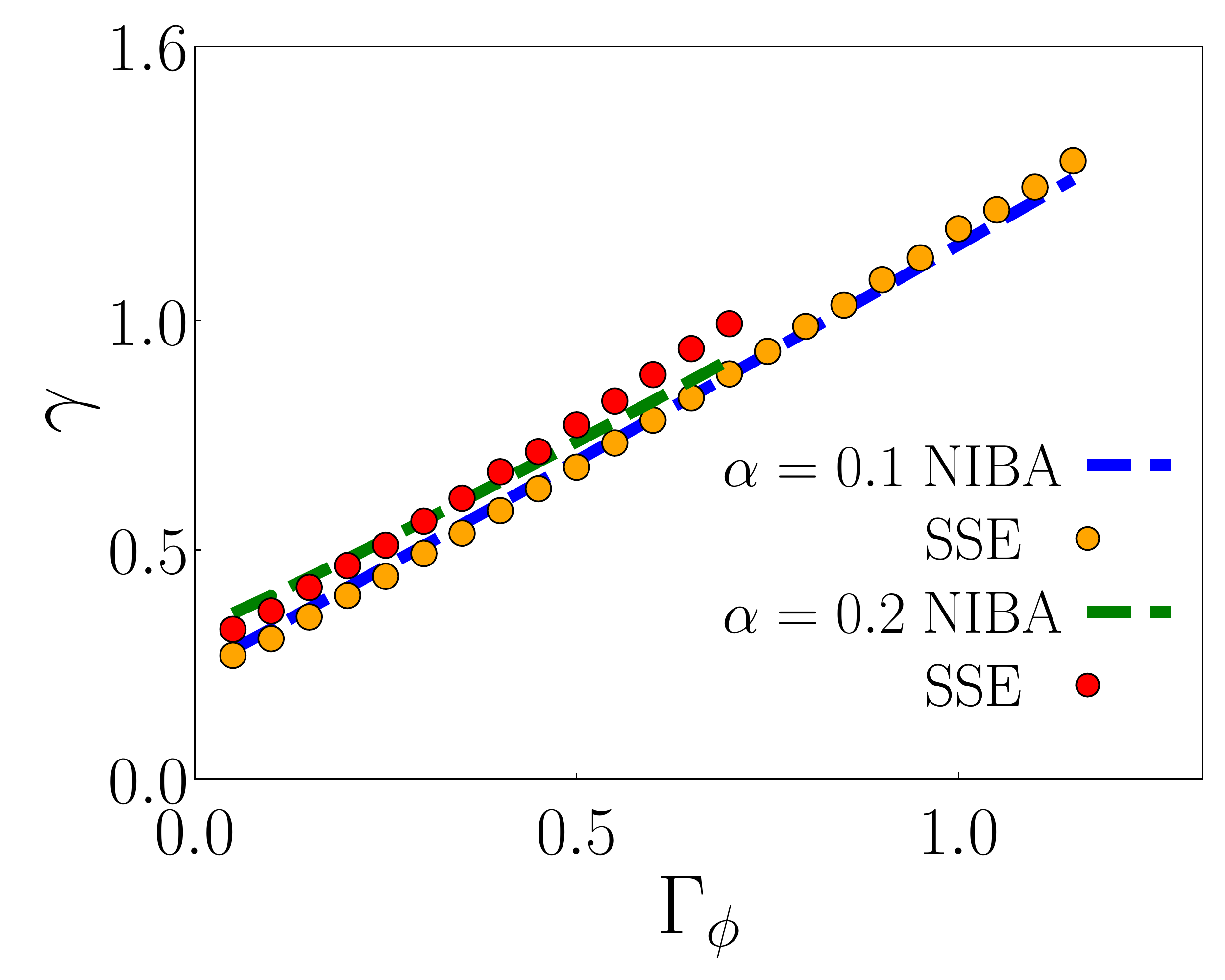}
\includegraphics [ scale=0.165]
{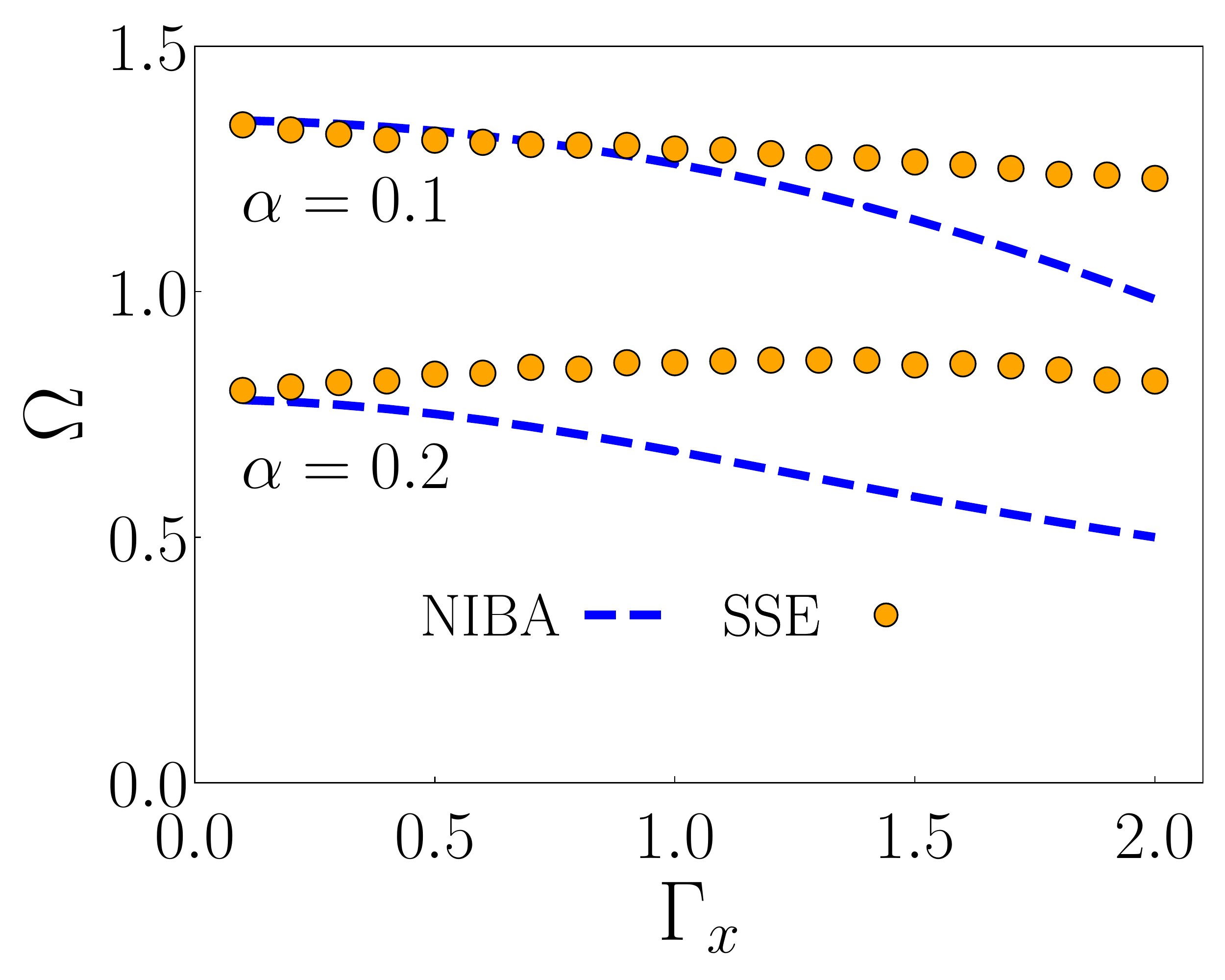}
\includegraphics [ scale=0.165]
{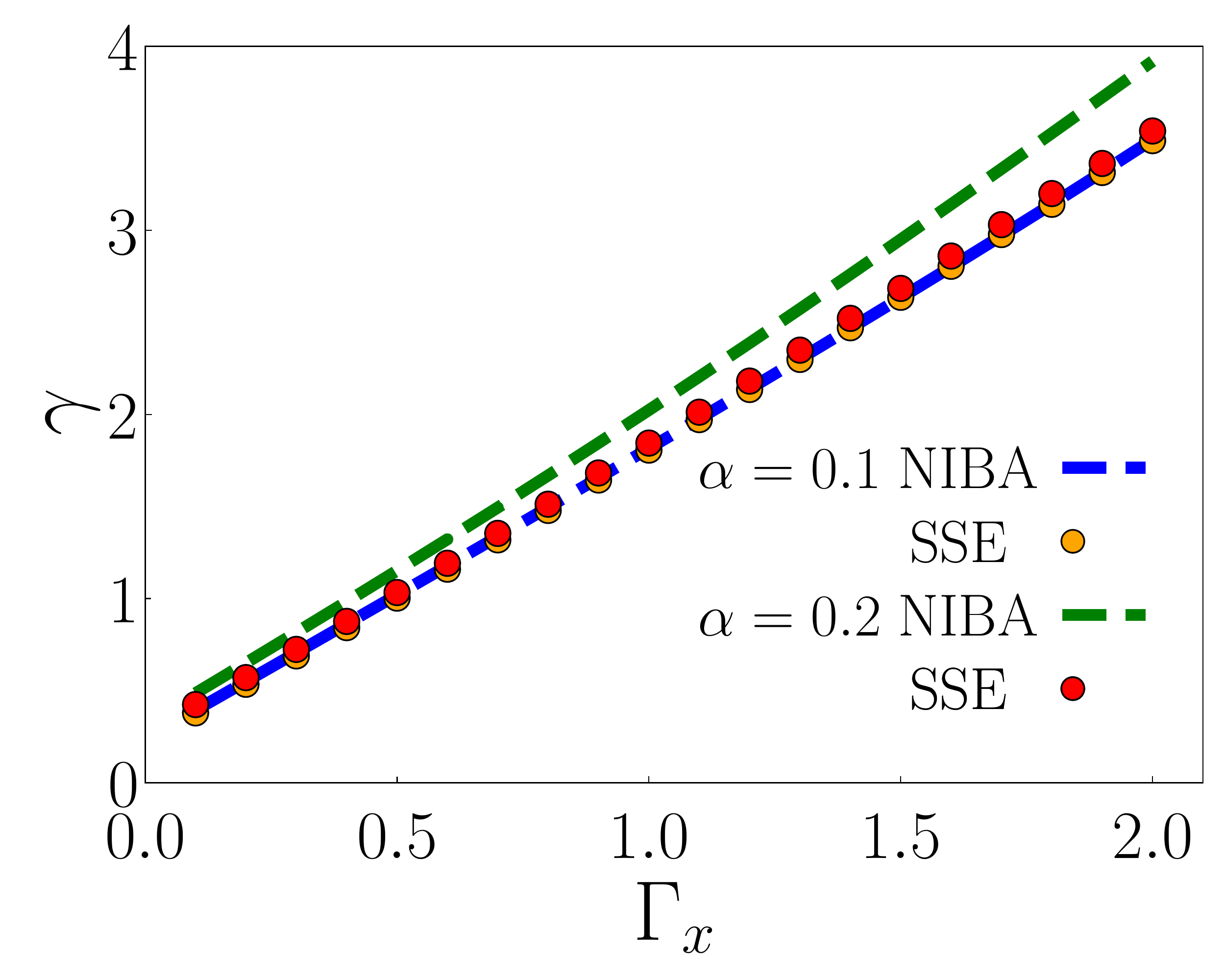}
\end{center} 
\caption{\label{fig:frequency_decay_rate} (color online)  Frequency $\Omega$ and decay rate $\gamma$ obtained from the SSE and NIBA as a function of the dissipation strength. The top panels depict $\Omega$ and $\gamma$ in the presence of dephasing. SSE and NIBA results are in excellent agreement. The frequency is very well described by the fit $\Omega=\sqrt{\Delta_{\rm eff}^2-\Gamma_\phi^2}$; see the dashed-dotted line.  
The bottom panels depict $\Omega$ and $\gamma$ in the presence of depolarization. The frequency $\Omega$ obtained from the SSE barely depends on $\Gamma_x$.}
\end{figure}

{To make even a more quantitative comparison between the SSE and NIBA, we also extract the frequency and the decay rate directly from the SSE by fitting the dynamics to the function $\langle\sigma_z(t)\rangle=A_0\cos(\Omega t+\phi_0) \exp(-\gamma t)$.}
Indeed, we find an excellent agreement between the exact SSE, the above fit and the NIBA in all cases involving dephasing.  
Extracting the decay rate $\gamma$ and frequency $\Omega$ from this fit, we show these parameters in Fig.~\ref{fig:frequency_decay_rate}, and contrast them with the prediction of the NIBA. 
Most notably, we observe a clear transition from underdamped to overdamped dynamics at sufficiently large $\Gamma_\phi$. In fact, we find that the dependence of the frequency $\Omega$ on $\Gamma_\phi$ is quantitatively consistent with the function 
\begin{equation}\label{eq:Omega_eff_dephasing}
\Omega\approx  \sqrt{\Omega_0^2 - \Gamma_\phi^2}\,.
\end{equation}
We recall that $\Omega_0$ is the oscillation frequency in the absence of Markovian dissipation (although it depends on $\alpha$). 
We can also fit the dissipation approximately to 
\begin{equation}\label{eq:gamma_eff_dephasing}
    \gamma\approx \gamma_0 + \Gamma_\phi,
\end{equation}
where $\gamma_0$ is again the effective decay rate in the absence of Markovian dissipation. 

Next we consider the depolarization channel. We refer the reader to \cref{App:NIBA} for details and just quote the final result from the NIBA:
\begin{equation} \label{eq:eq8blip_dph}
\begin{split}
\langle\sigma_z(t)\rangle&= \frac{e^{\lambda_1 t}(\lambda_1+\Gamma_x)}{(\lambda_1+\Gamma_x)+(1-2\alpha)(\lambda_1+2\Gamma_x)}\\&\quad+\frac{e^{\lambda_2 t}(\lambda_2+\Gamma_x)}{(\lambda_2+\Gamma_x)+(1-2\alpha)(\lambda_2+2\Gamma_x)}+P_{\rm inc},\\
\end{split}
\end{equation}
where $\lambda_{1,2}$ are the two solutions of 
\begin{equation}
(\lambda+2\Gamma_x)(\lambda+\Gamma_x)^{1-2\alpha}+\Delta_\text{\rm eff}^{2(1-\alpha)}=0\,. 
\end{equation}
In \cref{fig:magnetization_dephasing}, we show NIBA vs SSE in the presence of depolarization, and again find an excellent agreement. 
It can be shown that the last term in \cref{eq:eq8blip_dph} decays exponentially $P_{\rm inc}\sim \exp(-\Gamma_x t)$. This term must be included to ensure $\langle\sigma_z\rangle=1$ at $t=0$, but it can be ignored at intermediate or long times as it is suppressed exponentially. We remark that \cref{eq:eq8blip_dph} is derived in a  perturbative fashion in $\Gamma_x$. In this case too, the dynamics is generally characterized by a frequency $\Omega$ and an effective decay rate $\gamma$. Unlike the dephasing however, we find that the frequency changes only slightly with the rate of depolarization.  Indeed, the dynamics appears to be underdamped regardless of the depolarization rate, although it will decay more quickly for large dissipation (cf. \cref{fig:magnetization_SSE_Markovian}). For a quantitative comparison, we also plot the frequency $\Omega$ extracted from the SSE, and contrast that against the NIBA prediction in \cref{fig:frequency_decay_rate}. 
We find good agreement for $\Omega$ at small dissipation rate $\Gamma_x$ and an overall good agreement  for $\gamma$ in the entire range considered. We however observe that the NIBA prediction deviates from the exact SSE at larger values of $\Gamma_x$, which might be expected given the perturbative nature of \cref{eq:eq8blip_dph}. 
We can approximately fit the frequency and the decay rate as a function of $\Gamma_x$ as
\begin{equation}\label{eq:Omega_gamma_eff_depol}
\Omega \approx \Omega_0, \qquad \gamma \approx  \gamma_0 + 3\Gamma_x/2\,. 
\end{equation}

Before ending this section, an interpretation of its main results is in order. We recall that, in the absence of any Markovian dissipation, the spin $\langle\sigma_z(t)\rangle$ exhibits underdamped dynamics at the frequency $\Omega_0$ and with the decay rate $\gamma_0$, both depending nontrivially on the coupling to the Ohmic bath $\alpha$. As we turn on the Markovian dissipation, these characteristic frequency and time scales change in a simple fashion; cf.~\cref{eq:Omega_eff_dephasing,eq:gamma_eff_dephasing,eq:Omega_gamma_eff_depol}. 
The behavior for dephasing, for example, is reminiscent of an effective single-spin dynamics characterized by a Rabi frequency $\Omega_0$ and the  dephasing rate $\Gamma_\phi$. This would reproduce the frequency and the decay rate in \cref{eq:Omega_eff_dephasing,eq:gamma_eff_dephasing} assuming that $\gamma_0 \ll \Gamma_\phi$; however, this interpretation is not entirely correct since it doesn't capture the full aspects of the dynamics  \cite{HUR20082208}. 
Similarly, the behavior in \cref{eq:Omega_gamma_eff_depol} might suggest an effective dynamics characterized by depolarization and the Rabi frequency $\Omega_0$. This would be consistent with the fact that the frequency is almost independent of the depolarization rate. However, the effective depolarization rate would be $\Gamma_{\rm eff}= 3\Gamma_x/4$ which is smaller than the intrinsic dissipation rate $\Gamma_x$. Interestingly, this suggests that the depolarization rate is effectively \textit{reduced} due to the coupling to the bosonic bath. We emphasize again that this picture is incomplete since it would not correctly describe the dynamics more generally. In general, the coupling to the bosonic bath generates non-Markovian dynamics which cannot be mimicked by an effective Markovian master equation. 
\section{Summary and outlook}\label{sec:Outlook}
In this work, we have considered the dynamics of a single spin coupled to an Ohmic bath at zero temperature with the coupling strength $0<\alpha<1/2$ and a large cutoff $\omega_c$ as well as a Markovian bath inducing depolarization or dephasing. We have studied the dynamics using a nonperturbative approach known as the SSE valid in the regime of interest. Furthermore, we have derived analytic results based on the NIBA to gain insight into our exact numerical results. We have shown that NIBA is in excellent agreement with the SSE. 
Our results indicate that, under depolarization, the characteristic frequency of the spin oscillations is approximately unchanged from its renormalized value for a given $\alpha$, showing a kind of robustness against dissipation. On the other hand, dephasing changes the frequency, albeit in a simple fashion, and eventually renders the dynamics overdamped at large dissipation rates. 

The results obtained in this work hint at a simple picture where the renormalized spin interacts with the Markovian dissipation in a simple way. While an effective single-spin dynamics cannot describe the full dynamics, it is worthwhile finding an effective picture that consistently explains the main features. An interesting future direction is to identify to what extent the conclusions of this work extend to more general settings where $\alpha>1/2$ and the cutoff frequency is not restricted to large values. It is particularly interesting to determine the existence or destruction of the localization transition in the presence of Markovian dissipation \cite{dalla2010quantum,Torre_2012}. These questions will have practical consequences for the emergent quantum simulation platforms that aim to simulate spin-boson models and yet unavoidably come with intrinsic dissipation due to the unwanted coupling to the environment.  
\begin{acknowledgments}
%{\it Acknowledgments.}---% 
We thank Hassan Shapourian and Mohammad Mirhosseini for useful discussions. This work is supported by the Air Force Office of Scientific Research (AFOSR) under the award number FA9550-20-1-0073. We also acknowledge support from the National Science Foundation under the NSF CAREER Award (DMR2142866), as well as the NSF grants DMR1912799 and PHY2112893. 
\end{acknowledgments}
\appendix
\section{Stochastic Schr\"{o}dinger Equation}\label{sec:stochastic_sch}
We express the density matrix of the system plus bath in a Trotterized form as 
\begin{equation} \label{eq:eq0b4}
\begin{split}
|\rho(t)\rrangle&=\underbrace{e^{-i[H^u-H^l]\delta t}  ... e^{-i[H^u-H^l]\delta t}}_{n \ {times}}|\rho(0)\rrangle,
\end{split}
\end{equation}
where $n={t}/{\delta t}$ and $|\rho(0)\rrangle$ is initial density matrix of the system plus bath. We also assume that at time $t=0$ the system and the bath are decoupled, and take
\begin{align}
|\rho(0)\rrangle&=|\rho_S(0)\rrangle\otimes|e^{-\beta H_B}\rrangle,
\end{align}
where $\beta$ is the inverse temperature. At zero temperature, the density matrix of the bath is simply the vacuum. 
We then insert at the time slice $k$ in \cref{eq:eq0b4} the identity superoperator ${\cal I}_{SB}$ which acts on the system plus bath as 
\begin{equation}\label{eq:identity_Bath}
{\cal I}_{SB}=\sum_{\sigma_k^u,\sigma_k^l,N_k^u,N_k^l}|\sigma_k^u,\sigma_k^l, N_k^u,N_k^l\rangle\langle \sigma_k^u,\sigma_k^l, N_k^u,N_k^l|,
\end{equation}
where $|{N_k^{u/l}}\rangle$ and $|{\sigma_k^{u/l}}\rangle$ are the many-body complete basis of the bath and the system at time $k\delta t$ on the upper/lower branch of the Keldysh contour, respectively. The indices $|\sigma^{u/l}_k\rangle$ represent eigenstates of $\sigma_z^{u/l}$ while $|{N_k^{u/l}}\rangle$ could denote coherent states, the occupation number basis, or eigenstates of the bath Hamiltonian; the specific choice is unimportant. The system density matrix $|\rho_S(t)\rrangle$ is expressed  by tracing out the bath degrees of freedom as
\begin{align}\label{eq:trace}
|\rho_S(t)\rrangle&=\llangle I_B|\rho(t)\rrangle,
\end{align}
where the identity superket $|I_B\rrangle$ can be expressed as
\begin{align}
|I_{B}\rrangle=\sum_{N}| N N\rrangle.
\end{align}
By using \cref{eq:eq0b4,eq:identity_Bath,eq:trace}, the density matrix of the system in Trotterized form can be expressed as
\begin{align}
|\rho_S(t)\rrangle&=\sum_{\bar{\sigma},\bar{N}}\llangle I_B|e^{-i[H^u-H^l]\delta t}|\sigma_{n-1}^u\sigma_{n-1}^l N_{n-1}^uN_{n-1}^l \rangle\nonumber \\
&\times \langle \sigma_{n-1}^u\sigma_{n-1}^l N_{n-1}^uN_{n-1}^l|   \cdots |\sigma_2^u\sigma_2^l N_2^uN_2^l \rangle  \nonumber \\
&\times \langle \sigma_2^u\sigma_2^l N_2^uN_2^l |  e^{-i[H^u-H^l]\delta t}|\sigma_1^u\sigma_1^l N_1^uN_1^l \rangle \nonumber \\
&\times \langle \sigma_1^u\sigma_1^l N_1^uN_1^l |  e^{-i[H^u-H^l]\delta t}|\rho(0)\rrangle,
\end{align}
where $\bar\sigma$ (similarly $\bar N$) indicates the collections of all $\{\sigma^{u/l}_k\}$.
For $\delta t \rightarrow 0$, we can use the Trotter-Suzuki decomposition as  
\begin{align}\label{eq:Trotter}
e^{-i[H^u-H^l]\delta t}=e^{-i[H_S^u-H_S^l]\delta t}e^{-i[H_B^u+H_{SB}^u-H_B^l-H_{SB}^l]\delta t}e^{O({\delta t}^2)}.
\end{align}
With this identity, the density matrix of the system can be expressed as 
\begin{align}
\label{eq:eq0b4a1}
&|\rho_S(t)\rrangle=\sum_{\bar{\sigma},\bar{N}}\\
&\llangle I_B|e^{-i[H_S^u-H_S^l]\delta t} e^{-i[H^{'u}(\sigma^u_{n-1})-H^{'l}(\sigma_{n-1}^l)]\delta t}\nonumber
|\sigma_{n-1}^u\sigma_{n-1}^l N_{n-1}^uN_{n-1}^l \rangle \nonumber \\
&\times \langle \sigma_{n-1}^u\sigma_{n-1}^l N_{n-1}^uN_{n-1}^l|   \cdots \nonumber |\sigma_2^u\sigma_2^l N_2^uN_2^l \rangle \\
&\times \langle \sigma_2^u\sigma_2^l N_2^uN_2^l |e^{-i[H_S^u-H_S^l]\delta t}x e^{-i[H^{'u}(\sigma^u_{1})-H^{'l}(\sigma_{1}^l)]\delta t}  |\sigma_1^u\sigma_1^l N_1^uN_1^l \rangle \nonumber\\&\times \langle \sigma_1^u\sigma_1^l N_1^uN_1^l |e^{-i[H_S^u-H_S^l]\delta t} e^{-i[H^{'u}(\sigma^u_{0})-H^{'l}(\sigma_{0}^l)]\delta t}|\rho(0)\rrangle, \nonumber 
\end{align}
where $H^{'u/l}[\sigma^{u/l}(t)]$ are expressed as
\begin{equation}
H^{'u/l}[\sigma^{u/l}(t)]=H_B^{u/l}+\frac{1}{2}\sigma^{u/l}(t)\sum_k \lambda_k[b_k^{u/l}+{b_k^{\dagger}}^{u/l}].
\end{equation}
Finally, \cref{eq:trace} can be expressed as
\begin{align}
&|\rho_S(t)\rrangle=\lim_{\delta t\to 0}\sum_{\bar{\sigma}}e^{-i[H_S^u-H_S^l]\delta t}  |\sigma_{n-1}^u\sigma_{n-1}^l \rangle \langle \sigma_{n-1}^u\sigma_{n-1}^l |   ... \nonumber  \\
& |\sigma_2^u\sigma_2^l \rangle \langle \sigma_2^u\sigma_2^l |e^{-i[H_S^u-H_S^l]\delta t} |\sigma_1^u\sigma_1^l \rangle  \langle \sigma_1^u\sigma_1^l  |e^{-i[H_S^u-H_S^l]\delta t} |\rho_S(0)\rrangle \nonumber \\
& \times \llangle I_B| {\rm T}_t e^{-i \int_0^t ds(H^{'u}[\sigma^u(s)]-H^{'l}[\sigma^l(s)])}|e^{-\beta H_B}\rrangle. \label{eq:eq0b4b}
\end{align}
The last term of \cref{eq:eq0b4b} defines the influence of the bath on the system and is known as the influence functional $\phi[\sigma^{u/l}(t)]$:
\begin{equation} \label{eq:eq0b5}
\phi[\sigma^{u/l}(t)]= \llangle I_B|{\rm T}_te^{-i \int_0^t ds (H^{'u}[\sigma^u(s)]-H^{'l}[\sigma^l(s)])}|e^{-\beta H_B}\rrangle.
\end{equation}
To simplify \cref{eq:eq0b5}, we express $e^{-i \int_0^t ds(H^{'u}[\sigma^u(s)]-H^{'l}[\sigma^l(s)])}|e^{-\beta H_B}\rrangle$ in the operator form as 
\begin{equation} \label{eq:eq0b6}
\begin{split}
|\rho_B'(t)\rrangle&=e^{-i \int_0^t ds(H^{'u}[\sigma^u(s)]-H^{'l}[\sigma^l(s)])}|e^{-\beta H_B}\rrangle,\\
\frac{d|\rho_B'(t)\rrangle}{dt}&=-i (H^{'u}[\sigma^u(t)]-H^{'l}[\sigma^l(t)])|e^{-\beta H_B}\rrangle.
\end{split}
\end{equation}
We can again write the equation for vectorized $|\rho_B'(t)\rrangle$ in the operator matrix form as
\begin{equation} \label{eq:eq0b61}
\begin{split}
\frac{d\rho_B'(t)}{dt}&=-i[H'[\sigma^u(t)]e^{-\beta H_B}-e^{-\beta H_B}H'[\sigma^l(t)]],\\
\rho_B'(t)&=e^{-iH'[\sigma^u(t)]}e^{-\beta H_B}e^{iH'[\sigma^l(t)]}.
\end{split}
\end{equation}
Finally by using \cref{eq:eq0b5,eq:eq0b61}, the influence functional can be expressed as 
\begin{equation} \label{eq:eq0b7}
\begin{split}
\phi[\sigma^{u/l}(t)]&=\tr_B(e^{-iH'[\sigma^u(t)]}e^{-\beta H_B}e^{iH'[\sigma^l(t)]})\\
&=\tr_B(e^{-\beta H_B}e^{iH'[\sigma^l(t)]}e^{-iH'[\sigma^u(t)]}).
\end{split}
\end{equation}
This equation defines the general expression for the influence functional.
Since $H'$ is quadratic in terms of bosonic operators, it can be computed exactly to find 
\begin{equation} \label{eq:eq0b8}
\begin{split}
\phi[\xi(t),\eta(t)]&=\exp\Big[-\frac{1}{\pi}\int_{0}^{t} ds  \int_{0}^{s}ds'[-i L_{1}(s-s')\\&\times \xi(s)\eta(s')+ L_2(s-s')\xi(s)\xi(s')]\Big],
\end{split}
\end{equation}
where we have defined $\eta$ and $\xi$ as 
\begin{equation}
\begin{split}
\eta(s)&=\frac{\sigma^u(s)+\sigma^l(s)}{2},\\ 
\xi(s)&=\frac{\sigma^u(s)-\sigma^l(s)}{2}.
\end{split}
\end{equation}
The functions $L_2(s-s')$ and $L_1(s-s')$ are defined as the real and imaginary part of $\langle(\sum_k \lambda_k(b_k(s)+b_k^\dagger(s))) (\sum_k \lambda_k(b_k(s')+b_k^\dagger(s')))\rangle$, respectively, can be described in terms of the spectral function as
\begin{equation}
\begin{split}
L_1(t)&= \int_0^\infty d\omega J(\omega)\sin (\omega t),\\
L_2(t)&= \int_0^\infty d\omega J(\omega)\cos (\omega t)\coth(\beta\omega/2).
\end{split}
\end{equation}
Next, we consider that at time $t=0$, the system is in the $|\uparrow\rangle$ state in the $\sigma_z$ basis. 
Let us say that we are interested in the density matrix of the system at time $t$, and more specifically $\langle\uparrow|\rho(t)|\uparrow\rangle$, which in
 the vectorized form can be expressed as 
\begin{equation}\label{eq:density}
\langle\uparrow|\rho_s(t)|\uparrow\rangle=\llangle \uparrow\uparrow|\rho_s(t)\rrangle.
\end{equation}
By using \cref{eq:eq0b4b,eq:density} together with the definition of the influence functional can be written as 
\begin{align}
\label{eq:eq1}
 &\langle \uparrow |\rho_S(t)| \uparrow\rangle = \sum_{\overline{\sigma}_{u},\overline{\sigma}_{l}} \llangle \uparrow \uparrow| \cdots e^{-i\delta t \Delta[\sigma^{u}_{z}-\sigma^{l}_{z}]}|\sigma^{u}_{2}\sigma^{l}_{2}\rangle 
\nonumber\\&\times \langle \sigma^{u}_{2}\sigma^{l}_{2}| e^{-i\delta t \Delta[\sigma^{u}_{z}-\sigma^{l}_{z}]}|\sigma^{u}_{1}\sigma^{l}_{1}\rangle \langle \sigma^{u}_{1}\sigma^{l}_{1}|e^{-i\delta t \Delta[\sigma^{u}_{z}-\sigma^{l}_{z}]}|\uparrow \uparrow\rrangle \nonumber \\&\times \exp\Big[-\frac{1}{\pi}\int_{0}^{t} ds \int_{0}^{s}ds'(-i L_{1}(s-s')\xi(s)\eta(s')+\nonumber\\& \qquad\qquad + L_2(s-s')\xi(s)\xi(s'))\Big].
 \end{align}
For an Ohmic bath with the spectral function given by \cref{eq:eq0a1}, we have $L_1(s)=\pi^2\alpha \delta_\epsilon'(s)$ where we have defined $\delta_\epsilon(x) = \frac{1}{\pi}\frac{\epsilon}{x^+\epsilon^2}$, used the notation $\delta_\epsilon'(s)=d\delta_\epsilon(s)/ds$, and identified $\epsilon= 1/\omega_c$. We then compute the kernel corresponding to $L_1$ as
\begin{equation} \label{eq:eq2}
\begin{split}
&A_1(t)=\frac{i}{\pi}\int_{0}^{t} ds \int_{0}^{s}ds'L_{1}(s-s')\xi(s)\eta(s')\\
&=i\pi\alpha\int_{0}^{t} ds\xi(s)\Big[\eta(s')\delta_\epsilon(s-s')|_{0}^{s}- \int_{0}^{s}  ds'\frac{\partial \eta(s')}{\partial s'}\delta(s-s')\Big]\\
&=-i\pi\alpha\int_{0}^{t}ds\xi(s)\frac{\partial \eta(s)}{\partial s}\\
&\xrightarrow[\epsilon\to 0]{}  -i\pi\alpha\sum_{k} \xi(k)[\eta(k)-\eta(k-1)]
=i\pi\alpha\sum_{k} \xi(k)\eta(k-1),
 \end{split}
\end{equation}
where in the last line, we have taken the limit of $\epsilon \to 0$, which is justified for large $\omega_c$. We have also used the fact that $\xi(k)\eta(k)=0$. 
For the same Ohmic bath, we also find the kernel $L_2(s)=2\pi\alpha(1-\omega_c^2 s^2)/(1+\omega_c^2 s^2)^2$ which decays as a power law at long times. We can express $L_2(s)$ in a time window $[-t_{max}, t_{max}]$, where $t_{max}$ is the total simulation time beyond which $L_2(s)$ can be approximately taken to be zero. We now cast $L_2(s-s')$ in terms of a Fourier series so that $L_2$ becomes separable in time: 
\begin{equation} \label{eq:eq3}
\begin{split}
L_{2}(s-s')&= g_0+\sum_{m=1}^{m=m_{max}/2}g_m \cos\Big(\frac{m \pi(s-s')}{t_{max}}\Big).
\end{split}
\end{equation}
We then expand the cosine function as 
\begin{align}\label{eq:split1}
L_{2}(s-s')=&g_0+\sum_{m=1}^{m=m_{max}/2}g_m \Big(\cos\Big(\frac{m \pi s}{t_{max}}\Big) \cos\Big(\frac{m \pi s'}{t_{max}}\Big) \nonumber
\\& +\sin\Big(\frac{m \pi s}{t_{max}}\Big) \sin\Big(\frac{m \pi s'}{t_{max}}\Big)\Big),
\end{align}
where $g_0$ and $g_m$ are the Fourier series components given by
\begin{equation}
\begin{split}
g_0&=\frac{1}{2t_{max}}\int_{-t_{max}}^{t_{max}} ds L_2(s),\\
g_m&=\frac{1}{t_{max}}\int_{-t_{max}}^{t_{max}} ds L_2(s) \cos(m \pi s/t_{max}).
\end{split}
\end{equation}
We can recast \cref{eq:split1} as 
\begin{equation}\label{eq:split2}
\begin{split}
L_{2}(s-s')&=\sum_{m=0}^{m=m_{max}}G_m \psi_m(s)\psi_m(s').
 \end{split}
\end{equation}
By comparing \cref{eq:split1,eq:split2}, we obtain 
\begin{align}
G_0&=g_0,\quad G_{2m-1}=G_{2m}=g_m,\\
\psi_0(s)&=1,\quad \psi_{2m-1}(s)=\cos\Big(\frac{m \pi s}{t_{max}}\Big),  \psi_{2m}(s)=\sin\Big(\frac{m \pi s}{t_{max}}\Big) \nonumber .
\end{align}
Now the integral involving $L_2$ in \cref{eq:eq1} can be written as 
\begin{align} 
A_2(t)&=-\frac{1}{\pi}\int_{0}^{t}ds\int_{0}^{s} ds' L_2(s-s')\xi(s)\xi(s'),\nonumber \\
A_2(t)&=-\sum_{m=0}^{m=m_{max}} \frac{G_m}{\pi}\int_{0}^{t}ds\int_{0}^{s} ds' \xi(s)\xi(s')\psi_m(s)\psi_m(s'),\nonumber \\
&=-\frac{1}{2} \sum_{m=0}^{m=m_{max}}\frac{G_m}{\pi}\Big[\int_{0}^{t} ds\xi(s) \psi_m(s)\Big]^2. \label{eq:eq4}
\end{align}
Next, we introduce  the auxiliary fields $x_m$  corresponding to each frequency component $ (\omega_m)$ and employ the Hubbard-Stratonovich transformation as  \begin{equation} \label{eq:eq5}
\begin{split}
&e^{-\frac{1}{2}\sum_{m=0}^{m=m_{max}} \frac{G_m}{\pi}\Big[\int_{0}^{t} ds\xi(s)\psi_m(s)\Big]^2} \\
&=\prod_{m=0}^{m=m_{max}}\int \frac{dx_m}{\sqrt {2\pi}}  e^{-x_m^2/2}  e^{-i\int_{0}^{t}ds \xi(s)h(s)},  
\end{split}
\end{equation}
where we have defined $h(s)=\sum_{m=0}^{m=m_{max}} x_m \sqrt{\frac{G_m}{\pi}}\psi_m(s)$.
In a discretized time with the time step $\delta t$, we can now express \cref{eq:eq5} as 
\begin{equation} \label{eq:eq5a}
 e^{-i\int_{0}^{t}ds \xi(s)h(s)}=e^{-i\delta t\sum_{k} \xi(k)h(k)}.
\end{equation}
By substituting \cref{eq:eq2,eq:eq5a} into \cref{eq:eq1}, density matrix of the system is described as a sum over different spin configurations. 

Now we can see from \cref{eq:eq2,eq:eq5a} that the action of both $L_1$ as well as $L_2$ after the Hubbard-Stratonovich transformation becomes local in time involving only adjacent time steps. 
We can then express the matrix elements of the time-evolution generator $\tilde {\cal A}$ between two immediate time steps  as   
\begin{equation} \label{eq:eq12} 
\begin{split}
&\langle \sigma_{k}^u\sigma_{k}^l|\tilde {\cal A}|\sigma_{k-1}^u\sigma_{k-1}^l\rangle\\
&=e^{-i\delta t\xi(k)h(k)} e^{i\pi\alpha \eta({k-1})\xi(k)}\langle \sigma_{k}^u\sigma_{k}^l|e^{-i\delta t \frac{\Delta}{2} [\sigma_z^u-\sigma_z^l]}|\sigma_{k-1}^u\sigma_{k-1}^l\rangle \nonumber
\end{split}
\end{equation}
We can then write the matrix $\tilde {\cal A}$ as 
\begin{equation}
    \begin{split}
\tilde {\cal A}=
\begin{pmatrix}
1&i\delta t \frac{\Delta}{2} &-i\delta t \frac{\Delta}{2} &0\\
i\delta t\frac{\Delta}{2} e^{i\pi\alpha}&1-i\delta t h(k)&0&-i\delta t\frac{\Delta}{2} e^{-i\pi\alpha}\\
-i\delta t\frac{\Delta}{2} e^{-i\pi\alpha}&0&1+i\delta t h(k)&i\delta t\frac{\Delta}{2} e^{i\pi\alpha}\\
0&-i\delta t \frac{\Delta}{2} &i\delta t \frac{\Delta}{2} &1
\end{pmatrix}.
\end{split}
\end{equation}
Finally defining $\tilde {\cal A}= I-i\delta t {\cal A}(k)\approx 
 e^{-i\delta t {\cal A}(k)}$, we find 
\begin{equation} \label{eq:eq13}
\begin{split}
{\cal A}(k)=\begin{pmatrix}
0& \frac{\Delta}{2} &-\frac{\Delta}{2} &0\\
\frac{\Delta}{2} e^{i\pi\alpha}&- h(k)&0&-\frac{\Delta}{2} e^{-i\pi\alpha}\\
-\frac{\Delta}{2} e^{-i\pi\alpha}&0& h(k)&\frac{\Delta}{2} e^{i\pi\alpha}\\
0&- \frac{\Delta}{2} & \frac{\Delta}{2} &0
\end{pmatrix}
\end{split}
\end{equation}
and using \cref{eq:eq1,eq:eq2,eq:eq5,eq:eq12}, the density matrix $\rho_S(t)$ can be written as  
 \begin{align} \label{eq:eq14}
 \nonumber &\langle \uparrow|\rho_S(t)|\uparrow\rangle=\prod_{m=0}^{m=m_{max}} \int \frac{dx_m}{\sqrt {2\pi}} e^{-x_m^2/2} \llangle \uparrow\uparrow|{\rm T}_te^{-i\int_0^t ds {\cal A}(s)}|\uparrow\uparrow\rrangle\\
&\mbox{with} \quad {\cal A}(t)=
\begin{pmatrix}
0&-\frac{\Delta}{2}&\frac{\Delta}{2} &0\\
-\frac{\Delta}{2} e^{i\pi\alpha}&h(t)&0&\frac{\Delta}{2} e^{-i\pi\alpha}\\
\frac{\Delta}{2} e^{-i\pi\alpha}&0& -h(t)&-\frac{\Delta}{2} e^{i\pi\alpha}\\
0& \frac{\Delta}{2}&-\frac{\Delta}{2}&0
\end{pmatrix}. 
\end{align}
In order to solve \cref{eq:eq14}, we express ${\rm T}_te^{-i\int_0^t ds {\cal A}(s)}|\uparrow\uparrow\rrangle$ as
\begin{equation} \label{eq:eq14a}
\begin{split}
\frac{d|\psi(t)\rangle}{dt} = -i {\cal A}(t) |\psi(t)\rangle, 
\end{split}
\end{equation}
and, we recover \cref{eq:eq17a} in the main text.
\section{NIBA}\label{App:NIBA}
In this section, we extend the NIBA used in the  spin-boson model \cite{ Dekker_NIBA} to the case of  Markovian dissipation, specifically dephasing along both $z$ and $x$ directions.
\subsection{NIBA for depolarization}\label{subsec:NIBA_Dephasing}
We start with the time evolution of vectorized density matrix in the presence of depolarization (i.e., dephasing along the $x$ direction):
\begin{equation} \label{eq:eq0blip_dph}
\frac{d|\rho(t)\rrangle}{dt}={\cal L}|\rho(t)\rrangle,
\end{equation}
\begin{equation} \label{eq:eq1blip_dph}
{\cal L}=-i[H\otimes I-I\otimes H]+\Gamma_x(\sigma_x\otimes\sigma_x-I).
\end{equation}
We then follow the steps outlined in Ref.~\cite{ Dekker_NIBA}. 
We first apply a polaron transformation $U=\exp(-i\sigma_z B)\otimes \exp(-i\sigma_z B)$ with $B=i\sum_k (\lambda_k/\omega_k)(b_k^\dagger-b_k)$ on the Liouvillian ${\cal L}$:
\begin{equation} \label{eq:eq2blip_dph}
\tilde{\cal L}=-i[\tilde{H}\otimes I-I\otimes \tilde{H}]+\Gamma_x(\tilde{\sigma_x}\otimes\tilde{\sigma_x}-I),
\end{equation}
where $\tilde{H}=\Delta/2(\sigma_x \cos(B)-\sigma_y \sin(B))+\sum_k\omega_k b_k^\dagger b_k$ and $\tilde{\sigma_x}=\sigma_x \cos(B)-\sigma_y \sin(B)$.
We can write the equations of motion for $\sigma_x$, $\sigma_y$, and $\sigma_z$ as
\begin{equation} \label{eq:eq3blip_dph}
\begin{split}
\frac{d}{dt}\sigma_x&=-\Delta \sigma_z \sin B+\Gamma_x[\sigma_x\cos2B-\sigma_y\sin2B-\sigma_x],\\
\frac{d}{dt}\sigma_y&=-\Delta \sigma_z \cos B+\Gamma_x[-\sigma_y\cos 2B-\sigma_x\sin 2B-\sigma_y],\\
\frac{d}{dt}\sigma_z&=\Delta[\sigma_y\cos B+\sigma_x\sin B ]-2\Gamma_x\sigma_z.\\
 \end{split}
\end{equation}
From this equation,  $\sigma_x$ and $\sigma_y$ can be expressed as
\begin{equation} \label{eq:eq4blip_dph}
\begin{split}
\sigma_x(t) &=\int_0^t ds e^{-\Gamma_x (t-s)}\{-\Delta \sigma_z(s) \sin(B(s))\\&+\Gamma_x[\cos(2 B(s)) \sigma_x(s)-\cos(2 B(s)) \sigma_y(s)]\},
\\
\sigma_y(t) &=\int_0^t ds e^{-\Gamma_x (t-s)}\{-\Delta \sigma_z(s) \cos(B(s))\\&-\Gamma_x[\cos(2 B(s)) \sigma_y(s)-\sin(2 B(s)) \sigma_x(s)]\}.
 \end{split}
\end{equation}
We can now write the equation for $\sigma_z$ in \cref{eq:eq3blip_dph} as 
\begin{align}
\label{eq:eq5blip_dph}
&\frac{d}{dt}\sigma_z(t)+2\Gamma_x \sigma_z(t)\nonumber\\
=& \,\,\,\,\Delta \cos(B(t)) \int_0^t ds e^{\Gamma_x (s-t)}\{-\Delta \sigma_z(s) \cos(B(s))\nonumber\\& -\Gamma_x[\cos(2 B(s)) \sigma_y(s)-\sin(2 B(s)) \sigma_x(s)]\}\nonumber\\&
+\Delta \sin(B(t)) \int_0^t ds e^{\Gamma_x (s-t)}\{-\Delta \sigma_z(s) \sin(B(s))\nonumber\\&+\Gamma_x[\cos(2 B(s)) \sigma_x(s)-\cos(2 B(s)) \sigma_y(s)]\}
\end{align}
Finally, we take thermal average over both sides of \cref{eq:eq5blip_dph} and now assume that the spin and the bath are decoupled from each other. We then obtain
\begin{equation} \label{eq:eq6blip_dph}
\begin{split}
\bigg\langle \frac{d}{dt}\sigma_z(t)+2\Gamma_x \sigma_z(t)\bigg\rangle&=-\frac{\Delta^2}{2} \Big[\int_0^t ds e^{\Gamma_x(s-t)} \langle\sigma_z(s)\rangle \\&\langle e^{i B(t)} e^{-i B(s)}\rangle+ h.c\Big],\\
\bigg\langle \frac{d}{dt}\sigma_z(t)+2\Gamma_x \sigma_z(t)\bigg\rangle&=- \int_0^t ds e^{\Gamma_x(s-t)} \langle\sigma_z(s) f(t-s)\rangle.
\end{split}
\end{equation}
Notice that the terms involving $\sigma_x$ and $\sigma_y$ have disappeared because of the neutrality condition $\langle\exp(i (n B(t)-mB(s)))\rangle =0$ if $n\ne m$. In deriving the above equation, we have implicitly kept the terms to the first order in $\Gamma_x$ (assuming that the spin and the bath are decoupled to the first order in $\Gamma_x$). 
Now, we take the Laplace transform of the above equation to find
\begin{equation} \label{eq:eq7blip_dph}
\begin{split}
\langle\sigma_z(\lambda)\rangle&=\frac{1}{\lambda+2\Gamma_x+f(\lambda+\Gamma_x)}.
\end{split}
\end{equation}
where the Laplace transformed function $f(\lambda)$ is given by
\begin{align}
f(\lambda)&=\Delta_{\rm eff}\Big(\frac{\Delta_{\rm eff}}{\lambda}\Big)^{1-2\alpha},
\end{align}
and, for completeness, we recall the expression for $\Delta_{\rm eff}$:
\begin{align}
\Delta_{\rm eff}&=[\Gamma(1-2\alpha) \cos(\pi \alpha)]^{1/2(1-\alpha))}\Delta(\Delta/\omega_c)^{\alpha/(1-\alpha)}.
\end{align}

Finally, by taking inverse Laplace transform of \cref{eq:eq7blip_dph}, we find
\begin{equation}
\begin{split}
\langle\sigma_z(t)\rangle&= \frac{e^{\lambda_1 t}(\lambda_1+\Gamma_x)}{(\lambda_1+\Gamma_x)+(1-2\alpha)(\lambda_1+2\Gamma_x)}\\&+
\frac{e^{\lambda_2 t}(\lambda_2+\Gamma_x)}{(\lambda_2+\Gamma_x)+(1-2\alpha)(\lambda_2+2\Gamma_x)}+P_{\rm inc},
\end{split}
\end{equation}
where $P_{\rm inc}$ can be expressed as 
\begin{align}
P_{\rm inc}&=-\frac{1}{\pi}\int_0^\infty dz\frac{z^{1-2\alpha} e^{-(zt+\Gamma_x t)}\sin(\pi(1-2\alpha))\Delta_{\rm eff}^{2-2\alpha}}{D},\nonumber \\
D&=(-z+\Gamma_x)^2
z^{2(2-2\alpha)}+\Delta_{\rm eff}^{4-4\alpha}+ \\& 2\Delta_{\rm eff}^{2-2\alpha} (-z+\Gamma_x) z^{1-2\alpha}\cos(\pi(1-2\alpha)))\nonumber 
\end{align}
where $\lambda_{1,2}$ are the solutions to the equation $(\lambda+2\Gamma_x)(\lambda+\Gamma_x)^{1-2\alpha}+\Delta_{\rm eff}^{2(1-\alpha)}=0$.
\subsection{NIBA for dephasing}\label{subsec:NIBA_Depolarization}
In this section, we consider dephasing $\Gamma_\phi$ along $z$ direction.  By following similar steps to \cref{subsec:NIBA_Dephasing}, we find 
\begin{equation} \label{eq:eq3blip1}
\begin{split}
\frac{d}{dt}\sigma_x&=-\Delta\,\sigma_z \sin B-2\Gamma_\phi \sigma_x,\\
\frac{d}{dt}\sigma_y&=-\Delta \sigma_z \cos  B- 2\Gamma_\phi\sigma_y,\\
\frac{d}{dt}\sigma_z&= \Delta(\sigma_y\cos B+\sigma_x\sin B).\\
 \end{split}
\end{equation}
A similar procedure to \cref{subsec:NIBA_Dephasing} yields the equation of motion for $\langle\sigma_z\rangle$ as
\begin{equation} \label{eq:eq4blip}
\begin{split}
\frac{d}{dt}\langle\sigma_z\rangle&=-\int_0^t ds f(t-s)\langle\sigma_z(s)\rangle e^{2\Gamma_\phi(s-t)},\\
\langle\sigma_z(\lambda)\rangle&=\frac{1}{\lambda+f(\lambda+2\Gamma_\phi)}.
 \end{split}
\end{equation}
We point out that \cref{eq:eq4blip} is non-perturbative in $\Gamma_\phi$ unlike the case of depolarization.
The above equation can be now written as
\begin{equation}
\begin{split}
\langle\sigma_z(\lambda)\rangle&=\frac{(\lambda+2\Gamma_\phi)^{1-2\alpha}}{\lambda(\lambda+2\Gamma_\phi)^{1-2\alpha}+\Delta_{\rm eff}^{2-2\alpha}}.
 \end{split}
\end{equation}
Taking the inverse Laplace transform, we obtain 
\begin{equation} \label{eq:eq6blip}
\begin{split}
\langle\sigma_z(t)\rangle=\frac{e^{\lambda_1t}(\lambda_1+2\Gamma_\phi)}{2\lambda_1(1-\alpha)+2\Gamma_\phi}+\frac{e^{\lambda_2t}(\lambda_2+2\Gamma_\phi)}{2\lambda_2(1-\alpha)+2\Gamma_\phi}+P_{\rm inc},
\end{split}
\end{equation}
$P_{\rm inc}$ is expressed as 
\begin{equation}
P_{\rm inc}=-\int_0^\infty dy \sin(\pi-2\pi\alpha)/\pi e^{-y t} e^{-2\Gamma_\phi t} y^{(1-2\alpha)}\Delta_{\rm eff}^{(2-2\alpha)}/D_1,
\end{equation}
where $D_1$ is expressed as
\begin{align}
D_1&=((y+2\Gamma_\phi)(y+2\Gamma_\phi)y^{2-4\alpha}-2\cos(\pi-2\pi\alpha)\\&\Delta_{\rm eff}^{(2-2
\alpha)}(y+2\Gamma_\phi)y^{(1-2.0\alpha)}+ \Delta_{\rm eff}^{4-4\alpha}),
\end{align}
and $\lambda_{1,2}$ are given by the solutions to the equation $\lambda(\lambda+2\Gamma_\phi)^{1-2\alpha}+\Delta_{\rm eff}^{2-2\alpha}=0$.

\bibliography{Spin_Boson_ref}
%\appendix
\end{document}